\documentclass[final,5p,times,twocolumn,sort&compress]{elsarticle}
\usepackage{mathrsfs,amsmath,amsthm,latexsym,amssymb,amsfonts,epsfig,cancel,enumerate,graphicx,txfonts,subfigure}
\usepackage{multirow}
\numberwithin{equation}{section}
\usepackage{stfloats}
\usepackage{xcolor}
\definecolor{elsevierblue}{RGB}{10,144,205}
\AtBeginDocument{\hypersetup{colorlinks = true, linkcolor = elsevierblue, citecolor = elsevierblue, urlcolor = elsevierblue, filecolor = elsevierblue, }
}
\usepackage{etoolbox,orcidlink}
\usepackage{appendix}
\allowdisplaybreaks
\newcommand{\GZU}{School of Physics, Guizhou University, Guiyang 550025, China}
\journal{Physics Letters B}

\begin{document}

\begin{frontmatter}

\title{Image of a quantum-corrected black hole without Cauchy horizons illuminated by a static thin accretion disk}

\address[addr1]{\GZU}

\author[addr1]{Shilong Huang\orcidlink{0009-0002-5996-2952}}
\ead{gs.slhuang24@gzu.edu.cn}

\author[addr1]{Jiawei Chen\orcidlink{0009-0003-5390-8186}}
\ead{gs.chenjw23@gzu.edu.cn}

\author[addr1]{Jinsong Yang\orcidlink{0000-0003-4051-2767}\corref{cor1}}
\cortext[cor1]{Corresponding author.}
\ead{jsyang@gzu.edu.cn}

\begin{abstract}
Latest advances in effective quantum gravity propose a quantum-corrected black hole solution that avoids Cauchy horizons. This paper studies the images of this black hole when illuminated by a static thin accretion disk and explores the effect of the quantum parameter $\zeta$ on its appearance. First, we investigate the influence of $\zeta$ on the event horizon, photon sphere, critical impact parameter, and innermost stable circular orbit associated with the black hole. We find that all these quantities exhibit an increase with increasing $\zeta$. Meanwhile, we also use observational data from M87* and Sgr A* to impose constraints on $\zeta$ and compare the results with the theoretical constraint. Our analysis reveals that the observational constraint from Sgr A* is stronger than the theoretical one. We then derive the photon trajectory equation and analyze briefly the behavior of the trajectories. A detailed analysis shows that as $\zeta$ increases, the trajectories of photons undergo slight modifications when approaching the event horizon. Finally, by plotting the black hole's optical appearance under three emission models, we find that as $\zeta$ increases, the quantum-corrected black hole exhibits a larger shadow, along with narrower lensed and photon rings and reduced spacing between them. Furthermore, we also implement Johnson’s unbound distribution to simulate the image of the quantum-corrected black hole under large quantum parameters and reach the same conclusion. This work validates the rationality of this black hole solution through observational data, and provides its unique optical signatures that can serve as a promising avenue for probing quantum gravity effects near black holes.
\end{abstract}


\end{frontmatter}

\section{Introduction}

Shortly after its introduction, general relativity (GR) precisely explained the precession of Mercury's orbit~\cite{Einstein:1915bz} and successfully predicted gravitational redshift, light deflection, and gravitational waves~\cite{Einstein:1937qu,Peters:1964zz,Bondi:1962px}. These early achievements firmly established GR as a foundational theory of gravity, validating its ability to explain diverse astrophysical and cosmological observations. Subsequent research on light deflection expanded into studies of black hole (BH) shadows and optical appearances~\cite{Luminet:1979nyg,Holz:2002uf,Virbhadra:1999nm,Gralla:2019xty,Peng:2020wun,Hou:2021okc,Li:2021ypw,Hou:2022eev,Yang:2022btw,Zhang:2023okw,Zhang:2024lsf,He:2024amh,Chen:2025ifv,Chen:2025wut}. In 2015, gravitational waves--one of the key predictions of GR--were directly detected, providing compelling empirical evidence for the theory~\cite{LIGOScientific:2016aoc,LIGOScientific:2017vwq}. Then, in 2019, the Event Horizon Telescope (EHT) released the first image of the supermassive BH M87*, offering further direct confirmation of GR's validity in extreme gravitational fields~\cite{EventHorizonTelescope:2019dse}. This breakthrough significantly enhances the ability to probe BH environments and their intrinsic characteristics.

Although highly successful macroscopically, GR faces challenges such as the singularity problem~\cite{Penrose:1964wq,Hawking:1970zqf}, which leads to divergence and causality violation~\cite{Cassidy:1997kv}. As a fundamental limitation of GR, singularities indicate the breakdown of classical GR, necessitating new physics beyond it. To resolve these issues, physicists began modifying GR and incorporating quantum mechanics, leading to candidate quantum gravity theories~\cite{DeWitt:1967yk,Reuter:1996cp}. As a highly regarded quantum gravity theory, loop quantum gravity (LQG) aims to quantize spacetime geometry in a background-independent, non-perturbative framework~\cite{Rovelli:2004tv,Thiemann:2007pyv,Ashtekar:2004eh,Han:2005km}. To explore quantum gravity effects, LQG is applied to symmetry-reduced models, leading to loop quantum cosmology (LQC) and loop quantum black holes (LQBH). LQC resolves the Big Bang singularity by a quantum bounce~\cite{Ashtekar:2003hd,Bojowald:2005epg,Ashtekar:2006wn,Ding:2008tq,Yang:2009fp}. Meanwhile, LQBH aims to resolve BH singularities and to explore how quantum effects modify horizons, internal structures, and final states~\cite{Bojowald:2004af,Ashtekar:2005qt,Modesto:2008im,Gambini:2020nsf,Zhang:2020qxw,Zhang:2023yps}.

In effective theories arising from LQBH, the issue of covariance emerges. Recently, this issue has been successfully addressed by rigorously deriving the conditions for covariance in the spherically symmetric case~\cite{Zhang:2024khj,Zhang:2024ney}. Moreover, a fundamental connection between general covariance and Birkhoff's theorem has been revealed in~\cite{Zhang:2025ccx}. These developments have spawned a family of covariant quantum-corrected BH (QCBH) solutions, which enable to explore the phenomenological signatures of quantum geometry. Numerous studies have already been conducted with these solutions~\cite{Konoplya:2024lch,Liu:2024soc,Chen:2025ifv,Chen:2025aqh}. Preliminary studies of the optical appearance of the two QCBH solutions (referred to as QCBH-I and QCBH-II in Ref.~\cite{Chen:2025ifv}) have revealed different sensitivities of their observational features to the quantum correction mechanism: increasing the parameter $\zeta$ shrinks the shadow radius for the QCBH-I, whereas the optical appearance for the QCBH-II shows almost no alteration due to the quantum parameter~\cite{Chen:2025ifv}. This stark contrast highlights that quantum corrections can leave distinct and potentially observable features on BH images, motivating a systematic investigation across the entire solution family.

The recently proposed third QCBH solution is notable for its lack of Cauchy horizons. By eliminating the source of classical instabilities associated with inner horizons, it provides a more robust spacetime structure that avoids singularities~\cite{Zhang:2024ney}. It is worth noting that for this solution, a theoretical constraint on the quantum parameter $\zeta$ is naturally imposed to ensure the existence of an event horizon, resulting in $\zeta/M < 3.9374$. This raises a compelling question: whether this theoretical constraint is supported by observations. On the other hand, it also raises another fascinating phenomenological question: how these inherently stable quantum corrections manifest in the optical appearance of a BH when illuminated by an accretion disk.

To address these questions, we constrain the quantum parameter of this QCBH using available observational data. Meanwhile, to isolate the impact of quantum corrections on the appearance of BHs, we assume that this QCBH is surrounded by an optically thin and geometrically thin accretion disk (hereafter referred to as a thin accretion disk). In practice, realistic images arise from the complex interplay between the strong gravitational lensing of the BH and the electromagnetic plasma in the accretion flow, which requires intensive numerical general relativistic magnetohydrodynamic simulations~\cite{EventHorizonTelescope:2019pcy}. While more realistic accretion disk models may modify the specific morphology and intensity distribution of observed images~\cite{Vincent:2022fwj,Chael:2021rjo}, they also introduce additional factors that can obscure the direct signatures of quantum corrections. For instance, studies on optically thin and geometrically thick accretion disks in~\cite{Vincent:2022fwj} have found that photons are absorbed by the accretion disk itself when traversing a thicker disk medium, leading to a suppressed or even disappearance of photon ring signals. Although we adopt a simplified model, we believe that the qualitative trends in the main optical appearance changes (such as variations in shadow size, lensing-ring width, and photon-ring width) induced by the quantum parameter remain robust across different radiation models, with only quantitative differences.

The outline of this paper is as follows. We begin in Sec.~\ref{section2} by introducing the QCBH spacetime without Cauchy horizons, then utilize EHT observational data for M87* and Sgr A* to determine the allowable range of $\zeta$. In Sec.~\ref{section3}, we analyze photon trajectories and their dependence on $\zeta$. In Sec.~\ref{section4}, we examine the influence of the quantum parameter on different emission profiles from a thin accretion disk. Finally, we summarize the study in Sec.~\ref{section5}. In this paper, we work in units $G=c=1$.

\section{Quantum-corrected black hole and its photon sphere}\label{section2}

\begin{figure*}[htb]
	\centering
	\subfigure[$r_{\rm h}$]{\includegraphics[width=0.35\textwidth]{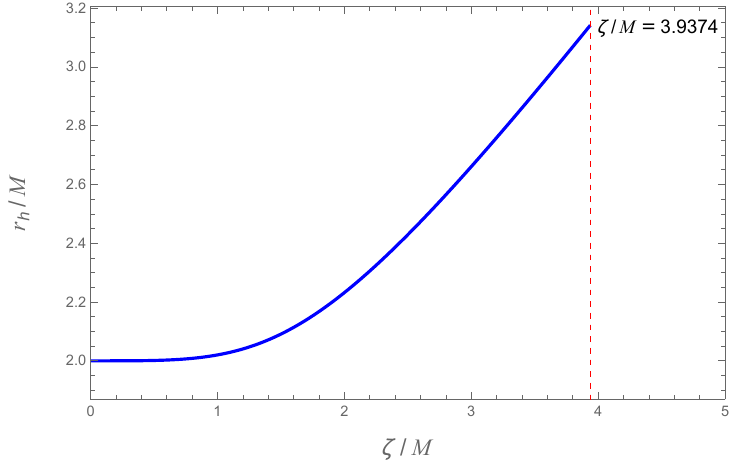}}
	\hspace{40pt}
	\subfigure[$r_{\rm ph}$]{\includegraphics[width=0.35\textwidth]{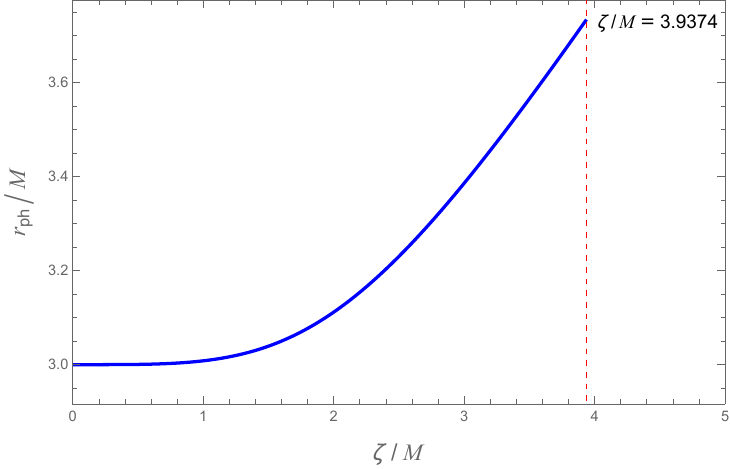}}

	\subfigure[$b_{\rm c}$]{\includegraphics[width=0.35\textwidth]{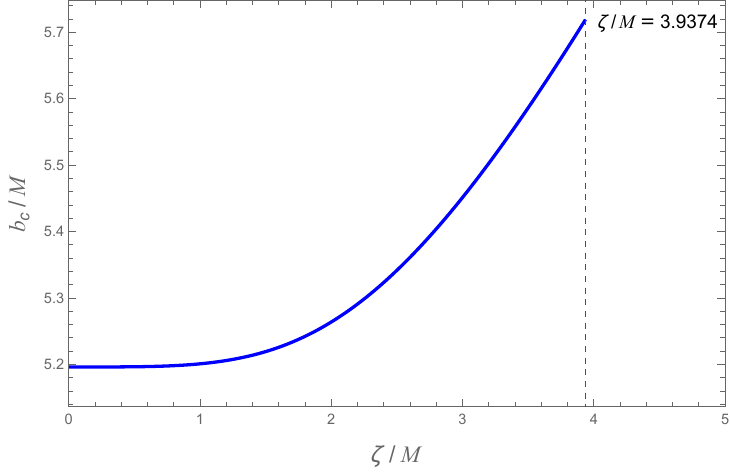}}
	\hspace{40pt}
	\subfigure[$r_{\rm isco}$]{\includegraphics[width=0.35\textwidth]{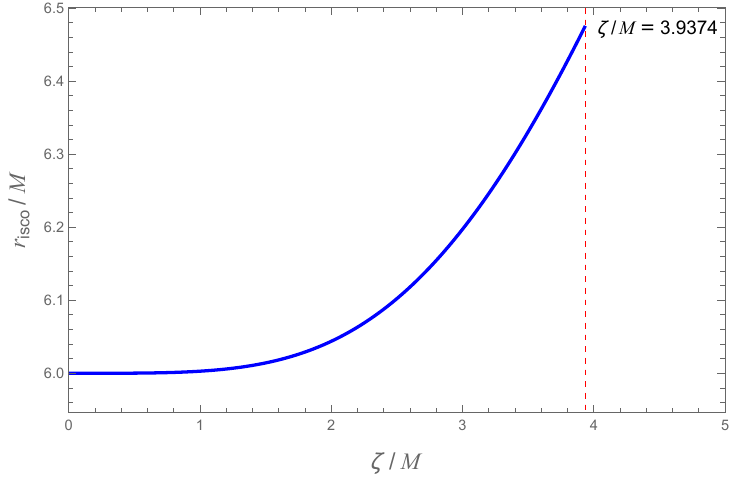}}

	\caption{The key physical quantities $r_{\rm h}$, $b_{\rm c}$, $r_{\rm ph}$, and $r_{\rm isco}$ vary with $\zeta/M$ in the QCBH. The vertical red dashed lines mark the location of the maximum allowed value $\zeta/M=3.9374$.}
	\label{fig:rh_bc_rph_risco}
\end{figure*}

Recently, a new spherically symmetric QCBH spacetime without Cauchy horizons that maintains general covariance was proposed in~\cite{Zhang:2024ney}. The line element of the spacetime reads
\begin{equation}
	{\rm d}s^{2} = - A(r){{\rm d}t}^{2} + \frac{1}{B(r)A(r)}{{\rm d}r}^{2} + C(r)({{\rm d}\theta}^{2} + \sin^{2}\theta{{\rm d}\varphi}^{2}),\label{line_element}
\end{equation}
where
\begin{equation}
	\begin{split}
		A(r) &= 1 - (-1)^{n}\frac{r^{2}}{\zeta^{2}}\arcsin\left(\frac{2M\zeta^{2}}{r^{3}}\right) - \frac{n\pi r^{2}}{\zeta^{2}},\\
		B(r)&= 1 - \frac{4M^{2}\zeta^{4}}{r^{6}},\\
		C(r)&= r^{2}.\label{metric_compoent}
	\end{split}
\end{equation}
Here $M$ stands for the ADM mass, $n$ is an arbitrary integer, and $\zeta$ denotes the quantum parameter. A straightforward calculation shows that the metric reduces to the Schwarzschild metric when $n = 0$ and $\zeta$ approaches zero. In this paper, we focus on the $n = 0$ case and take $\zeta = 0$ to represent the Schwarzschild case for simplicity of presentation. The physical quantities ($r$, $\zeta$, $M$) are converted into the dimensionless quantities ($r/M$, $\zeta/M$, 1).

The domain of the arcsin function requires
\begin{equation}
	r \geq (2M\zeta^{2})^{1/3} \equiv r_{\rm min}.\label{r_max}
\end{equation}
Within this domain, there exist three different spacetime structures depending on the value of $M$, each of which contains a BH region, as classified in Ref.~\cite{Zhang:2024ney}. For a BH with an event horizon, $A(r)$ must have a real root, which occurs when
\begin{align}
	M>\frac{\zeta}{2}\left( \frac{2}{\pi} \right)^{3/2},\label{eq:Mvalue}
\end{align}
and we denote this root by $r = r_{\rm h}$ (its dependence on $\zeta/M$ is shown in the first panel of Fig.~\ref{fig:rh_bc_rph_risco}). Equation~\eqref{eq:Mvalue} implies a theoretical bound on $\zeta/M$
\begin{equation}
	\frac{\zeta}{M} < 2\left( \frac{\pi}{2} \right)^{3/2} \cong 3.9374.\label{zeta_limit}
\end{equation}
In the present paper, we only consider the BH with $M$ satisfying Eq.~\eqref{eq:Mvalue}.

Based on the spherical symmetry of the spacetime, there exist two Killing vector fields $\left(\frac{\partial}{\partial t}\right)^{a}$ and $\left(\frac{\partial}{\partial\varphi}\right)^{a}$. For a geodesic with affine parameter $\tau$ and tangent vector $\left(\frac{\partial}{\partial\tau}\right)^{a}$, its motion can always be confined to the equatorial plane by adopting Schwarzschild coordinates with $\theta=\pi/2$. The inner products of the tangent vector with these two Killing vector fields yield two conserved quantities along the geodesic, denoted as $E$ and $L$~\cite{Wald:1984rg,Liang:2023ahd}

\begin{equation}
	\begin{split}
		&E:=-g_{a b}\left(\frac{\partial}{\partial t}\right)^{a}\left(\frac{\partial}{\partial \tau}\right)^{b}=A(r)\dot{t},\\
		&L:=g_{a b}\left(\frac{\partial}{\partial \varphi}\right)^{a}\left(\frac{\partial}{\partial \tau}\right)^{b}=C(r) \dot{\varphi},
	\end{split}\label{energy_and_angular_monmentum}
\end{equation}
which represent the energy and angular momentum of the photon or massive particle, respectively, and the dot denotes differentiation with respect to $\tau$. Furthermore, we have~\cite{Wald:1984rg,Liang:2023ahd}
\begin{equation}
	- \kappa = g_{ab}\left( \frac{\partial}{\partial\tau} \right)^{a}\left( \frac{\partial}{\partial\tau} \right)^{b},\label{kappa_definition}
\end{equation}
where
\begin{equation}
	\kappa =
	\begin{cases}
		1, & \text{for timelike geodesic.} \\
		0, & \text{for null geodesic.}
	\end{cases}\label{kappa_values}
\end{equation}
We are interested in the photon trajectories ($\kappa =0$) near the QCBH described by Eq.~\eqref{line_element}. Then, we obtain
\begin{equation}
		\dot{r}^{2} = \left(E^{2} - V_{\rm eff}(r)L^2 \right)B(r),\label{r_dot}
\end{equation}
where $V_{\rm eff}(r)$ denotes the photon's effective potential, defined as
	\begin{equation}
		V_{\rm eff}(r) = \frac{A(r)}{C(r)}=\frac{1}{r^2} - \frac{\arcsin\left( \frac{2 M \zeta^2}{r^3} \right)}{\zeta^2}.\label{effective_potential}
	\end{equation}
By introducing the impact parameter $b \equiv\frac{L}{E}$, and combining Eqs.~\eqref{energy_and_angular_monmentum} and \eqref{r_dot}, we derive the photon trajectory equation as
\begin{equation}
	\left( \frac{{\rm d}r}{{\rm d}\varphi} \right)^{2} = {C(r)}^{2}\left( \frac{1}{b^{2}} - V_{\rm eff}(r) \right)B(r).\label{trajectory_equation}
\end{equation}
\begin{figure}[htbp]
	\centering
	\includegraphics[width=7cm]{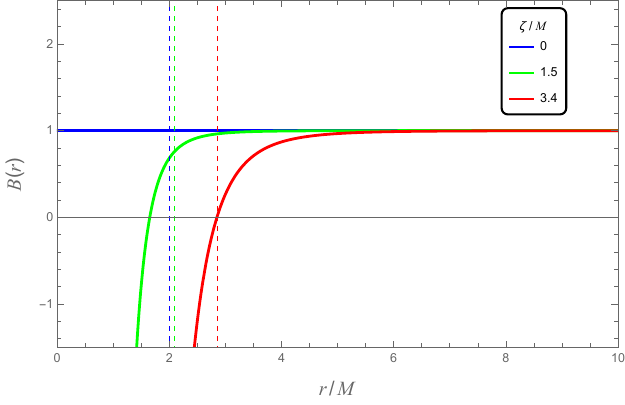}

	\caption{Variation of $B(r)$ with radial coordinate $r$ at three $\zeta/M$ values: 0 (blue), 1.5 (green), and 3.4 (red). The dashed lines mark where the event horizon forms for each $\zeta/M$.}
	\label{fig:function_μ(r)}
\end{figure}
\begin{figure}[htbp]
	\centering
	\includegraphics[width=7cm]{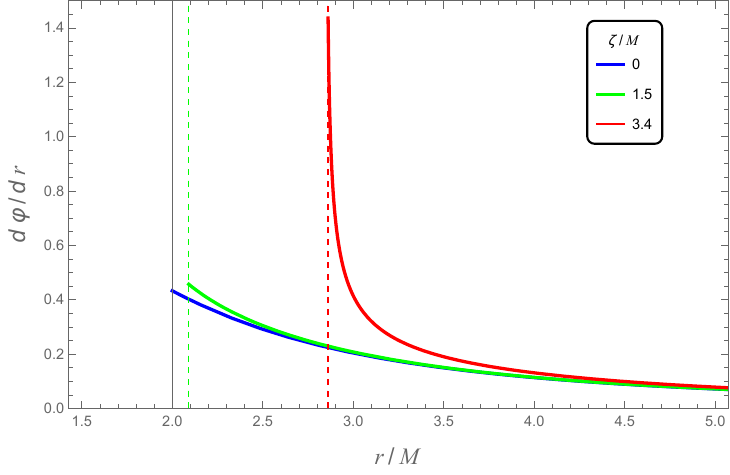}

	\caption{Variation of the deflection rate $\frac{{\rm d}\varphi}{{\rm d}r}$ with $r$ for selected values of $\zeta/M$: 0 (blue), 1.5 (green), and 3.4 (red). The dashed vertical lines indicate the event horizon for each case. The impact parameter is fixed at $b=b_{\rm c}/3$.}
	\label{fig:deflection_rate_dφdr}
\end{figure}

\begin{figure*}[htb]
	\begin{minipage}{0.45\textwidth}
		\includegraphics[width=3in,height=5.5in,keepaspectratio]{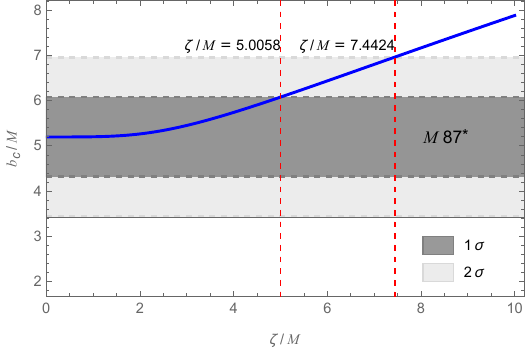}
	\end{minipage}
	\hfill
	\begin{minipage}{0.45\textwidth}
		\includegraphics[width=3in,height=5.5in,keepaspectratio]{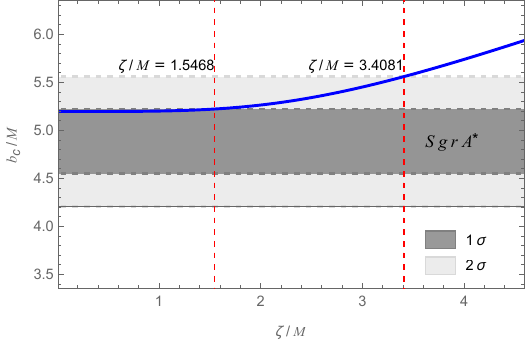}
	\end{minipage}

	\caption{Constraints on $\zeta/M$. The left panel corresponds to M87* data, while the right one to Sgr A* data. The dark gray regions indicate the BHs size ranges at the $1\sigma$ confidence level, and the light gray regions indicate the ranges at the $2\sigma$ confidence level. The two red dashed lines mark the maximum allowed $\zeta/M$ values under the different confidence levels.}
	\label{fig:constraint}
\end{figure*}

The main difference between trajectory equation~\eqref{trajectory_equation} in the QCBH and that in the Schwarzschild BH is the additional term $B(r)$, which modifies the radial metric by quantum geometry. Figure~\ref{fig:function_μ(r)} illustrates the behavior of the function $B(r)$ at selected values of $\zeta/M$. We can see that when $\zeta/M$ is large, $B(r)$ decreases rapidly near the event horizon, which may affect photon trajectories. This effect becomes more evident in Fig.~\ref{fig:deflection_rate_dφdr}, which plots $\frac{{\rm d}\varphi}{{\rm d}r}$. At $\zeta/M=3.4$, the rate of change of $\varphi$ with $r$ is significantly greater than 1, indicating that photons may undergo special deflection near the event horizon due to the influence of the quantum parameter.

Among the trajectories of photons orbiting the BH, there exists a special type of photon orbit, namely, the unstable photon sphere orbit. Photons in this orbit, when perturbed, either fall into the BH or moves towards infinity. The behavior of these orbits is governed by the properties of the effective potential. To identify the photon sphere radius and its instability, the following criteria must be met:
\begin{enumerate}
\item The first derivative of the effective potential vanishes at the photon sphere radius
\begin{equation}
	\left. \frac{{\rm d}V_{\rm eff}(r)}{{\rm d}r} \right|_{r = r_{\rm ph}} = 0,\label{eq:first_condition}
\end{equation}
indicating an extremum point in the potential. Within the domain of positive real numbers, we can find a solution expressed as
	\begin{equation}
		r_{\rm ph} =\sqrt{3 M^2 + \frac{9 M^4}{u} + u},\label{eq:rph}
	\end{equation}
	where $u = \left(27 M^6 + 2M^2 \zeta^4 + 2M^2 \zeta^2 \sqrt{27 M^4 +\zeta^4}\right)^{1/3}$.

\item The second derivative of the effective potential at the photon sphere radius is negative.
	\begin{equation}
		\left. \frac{{\rm d}^{2}V_{\rm eff}(r)}{{{\rm d}r}^{2}} \right|_{r = r_{\rm ph}} < 0,\label{eq:second_condition}
	\end{equation}
confirming that the extremum is a local maximum and thus unstable.
\end{enumerate}

Additionally, the critical impact parameter $b_{\rm c}$ is defined through the value of the effective potential at the photon sphere, namely
\begin{equation}
	V_{\rm eff}(r) \bigg|_{r = r_{\rm ph}} = \frac{1}{b_{\rm c}^{2}}.\label{eq:third_condition}
\end{equation}
This expression relates the photon sphere radius to the impact parameter of photons moving along unstable circular paths. These conditions collectively characterize the photon sphere and its instability. Substituting Eqs.~\eqref{eq:rph} and \eqref{effective_potential} into Eq.~\eqref{eq:third_condition}, we obtain the expression for $b_{\rm c}$ as
	\begin{equation}
		b_{\rm c} =\frac{1}{\sqrt{\frac{1}{3 M^2 + \frac{9 M^4}{u} + u} - \frac{\arcsin\left( \frac{2 M \zeta^2}{\left(3 M^2 + \frac{9 M^4}{u} + u\right)^{3/2}} \right)}{\zeta^2}}}.\label{eq:bc}
	\end{equation}
In panels (b) and (c) of Fig.~\ref{fig:rh_bc_rph_risco}, we show the dependence of both $r_{\rm ph}$ and $b_{\rm c}$ on $\zeta/M$. Their behavior exhibit pattern consistent with those of $r_{\rm h}$.

For a massive particle ($\kappa =1$), its innermost stable circular orbit (isco) satisfies the following expression~\cite{Chen:2025ifv,Wang:2023vcv}:
\begin{equation}
	2B(r_{\rm isco})A(r_{\rm isco}) \left[r_{\rm isco}\left(\frac{2{A}'(r_{\rm isco})}{A(r_{\rm isco})}-\frac{{A}''(r_{\rm isco})}{{A}'(r_{\rm isco})}\right)-3 \right]=0,\label{isco}
\end{equation}
where the prime denotes differentiation with respect to $r$. Outside the event horizon, $A(r)>0$ and $0 < B(r) \leq 1$, thus we obtain:
\begin{equation}
	r_{\rm isco} = \frac{3 A(r_{\rm isco}) {A}'(r_{\rm isco})}
	{2 {A}'(r_{\rm isco})^2 - A(r_{\rm isco}) {A}''(r_{\rm isco})}.
	\label{risco}
\end{equation}

The panel (d) in Fig.~\ref{fig:rh_bc_rph_risco} shows the beharvior of $r_{\rm isco}$ as the function of $\zeta/M$. We can see from Fig.~\ref{fig:rh_bc_rph_risco} that these four physical quantities ($r_{\rm h}$, $b_{\rm c}$, $r_{\rm ph}$, $r_{\rm isco}$) all increase monotonically with $\zeta/M$.

A theoretical upper bound on $\zeta/M$ has been established based on the existence of the BH event horizon in Eq.~\eqref{zeta_limit} earlier. To verify whether the theoretical constraint is consistent with the observational constraints, and to investigate the effect of the quantum parameter on the images of the QCBH within a meaningful parameter range, it is necessary to compare the model with astronomical observational data. The EHT measurements of shadow sizes for the supermassive BHs M87* and Sgr A* offer a direct and powerful tool for constraining parameters of various BH models~\cite{EventHorizonTelescope:2021dqv,EventHorizonTelescope:2022wkp}. Although M87* and Sgr A* are widely regarded as rotating Kerr BHs according to GR, it is reasonable to employ their shadow size measurements to constrain a spherically symmetric QCBH, based on the following considerations. Firstly, for Sgr A*, the relativistic precession of stellar orbits suggests a low spin ($a_{*}\lesssim 0.1$) under certain assumptions about the orbital geometry~\cite{Fragione:2020khu,Fragione:2022oau}. In such a low-spin regime, the impact of spin on the shadow size is negligible~\cite{Gralla:2020srx,EventHorizonTelescope:2021dqv}. Secondly, for M87*, despite strong evidence supporting the hypothesis that the central object is a Kerr BH~\cite{EventHorizonTelescope:2019ggy}, the limited resolution of the M87* image renders the Schwarzschild solution a conservative yet viable choice within the allowed observational uncertainties~\cite{EventHorizonTelescope:2021dqv}. Finally, it is crucial to note that while the EHT's posterior distribution for the Kerr BH angular gravitational radius is derived from Kerr BH simulations~\cite{EventHorizonTelescope:2019ggy,EventHorizonTelescope:2022wkp}, as argued in~\cite{Mizuno:2018lxz}, given the current observational quality, images generated from non-Kerr simulations would be largely indistinguishable from those of Kerr BH simulations. Consequently, the EHT shadow observations remain applicable for constraining BH parameters across various gravitational theories, and more precise observational data will aid in distinguishing among different gravitational theories~\cite{Cunha:2019ikd,Bambi:2019tjh,Khodadi:2020jij,Ghosh:2020spb,Afrin:2021wlj,Khodadi:2021gbc,Afrin:2021imp,Kuang:2022ojj}. Thus, the use of their observational data to constrain the quantum parameter of the QCBH remains valid.

\begin{table}[htbp]\centering

	\caption{BH shadow size at different confidence levels.}
	\setlength{\tabcolsep}{3pt}
	\begin{tabular}{ccccc}
		\hline
		BH & $1\sigma$ & $2\sigma$ & Reference \\
		\hline
		M87* & $4.313\leq b_{\rm c}\leq 6.080$ &$3.430\leq b_{\rm c}\leq 6.963$ &~\cite{EventHorizonTelescope:2021dqv} \\
		Sgr A* & $4.547\leq b_{\rm c}\leq 5.222$ & $4.209\leq b_{\rm c}\leq 5.560$ &~\cite{EventHorizonTelescope:2022wkp}
		\label{table:observation}\\
		\hline
	\end{tabular}
\end{table}

Following the established methodology of EHT analyses, we directly use the ranges of the critical impact parameter $b_{\rm c}$ at the $1\sigma$ and $2\sigma$ confidence levels to perform the constraints, as summarized in Table~\ref{table:observation}. By comparing the predicted values of $b_{\rm c}$ across a range of $\zeta/M$ with these observationally allowed regions (see Fig.~\ref{fig:constraint}), we obtain corresponding limits on the quantum parameter: at the $1\sigma$ level, $\zeta/M \leq 5.0058$ (from M87*) or $\zeta/M \leq 1.5468$ (from Sgr A*); at the more conservative $2\sigma$ level, the limits relax to $\zeta/M \leq 7.4424$ (M87*) or $\zeta/M \leq 3.4081$ (Sgr A*). The reliability of the data for M87* and Sgr A* has been discussed in~\cite{EventHorizonTelescope:2021dqv,Kuang:2022ojj,EventHorizonTelescope:2022wkp}; here, we select the more reliable data from Sgr A*. Its $2\sigma$ upper limit ($\zeta/M \leq 3.4081$) is particularly noteworthy. It not only accommodates greater measurement uncertainty but also exhibits a striking proximity to the theoretical ceiling from Eq.~\eqref{zeta_limit}. The consistency between this conservative observational constraint and the theoretical bound significantly strengthens the physical plausibility of the parameter space we explore. Therefore, we adopt $\zeta/M \leq 3.4081$ as our fiducial observational limit for the subsequent investigation of the model's optical signatures, ensuring that our study is anchored in both theoretical coherence and empirical consistency.

\begin{table*}[hb]

	\caption{Theoretical quantities for the Schwarzschild BH and QCBH, computed at different $\zeta/M$.}
	\setlength{\tabcolsep}{10.5pt}
	\begin{tabular}{cccccccccc}
		\hline
		$\zeta/M$&$r_{\rm h}/M$&$r_{\rm ph}/M$&$r_{\rm isco}/M$&$b_{\rm c}/M$&$b_1^-/M$ &$b_2^-/M$ &$b_2^+/M$ &$b_3^-/M$ &$b_3^+/M$ \\
		\hline
		0 & 2 & 3 & 6 & 5.19615 & 2.84770 & 5.01514 & 6.16757 & 5.18781 & 5.22794\\
		1.5 & 2.09107 & 3.03930 & 6.01417 & 5.21925 & 2.91010 & 5.05516 & 6.17283 & 5.21229 & 5.24839\\
		2.5 & 2.42931 & 3.32031 & 6.10218 & 5.34161 & 3.13444 & 5.22441 & 6.20688 & 5.33805 & 5.36112\\
		3.4 & 2.86078 & 3.52832 & 6.30128 & 5.55759 & 3.40584 & 5.47389 & 6.29139 & 5.55584 & 5.56901 \label{table:1}\\
		\hline
	\end{tabular}
\end{table*}

\section{Photon trajectories}\label{section3}

\begin{figure}[bp]
	\centering
	\includegraphics[width=7.5cm]{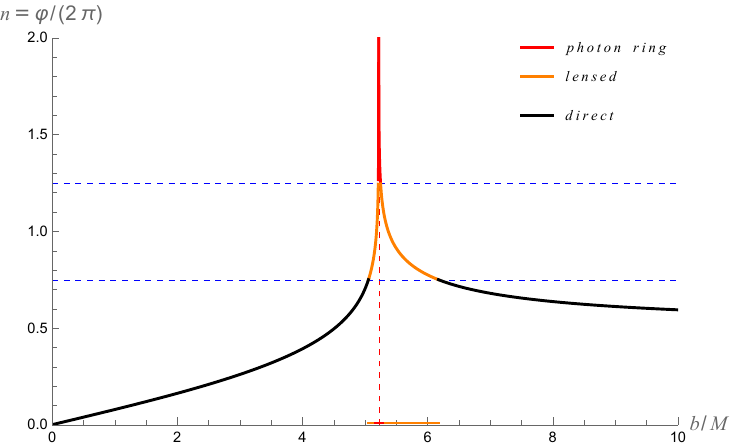}

	\caption{Total number of orbits with $\zeta/M=1.5$.}
	\label{fig:total_number_of_orbit}
\end{figure}
\begin{figure*}[hb]
	\centering
	\subfigure[$\zeta/M =0$]{\includegraphics[width=0.32\textwidth]{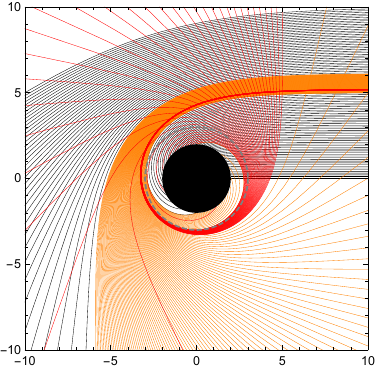}}
	\hspace{35pt}
	\subfigure[$\zeta/M =1.5$]{\includegraphics[width=0.32\textwidth]{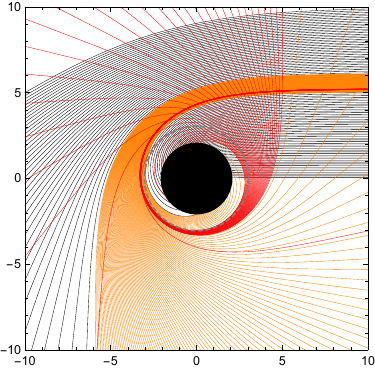}}

	\subfigure[$\zeta/M =2.5$]{\includegraphics[width=0.32\textwidth]{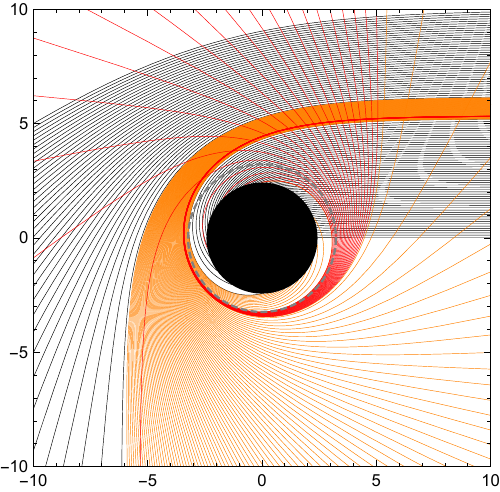}}
	\hspace{35pt}
	\subfigure[$\zeta/M =3.4$]{\includegraphics[width=0.32\textwidth]{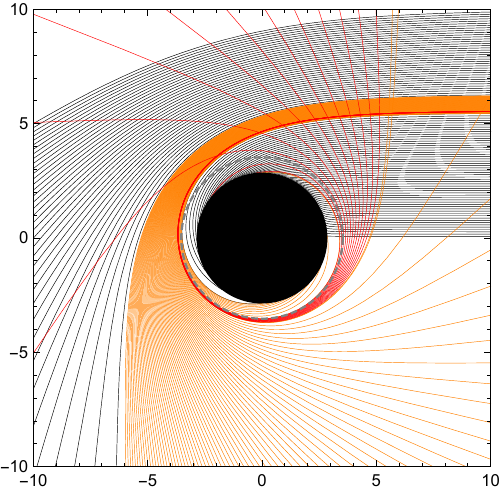}}

	\caption{Photon trajectories plotted against the impact parameter $b$ for selected quantum parameters. The red, orange, and black lines stand for the photon ring, lensed ring, and direct trajectories. A black disk shows the BH, and a dashed black curve marks the photon sphere.}
	\label{fig:photon_trajectories}
\end{figure*}
In this section, we employ the ray-tracing technique to study the light trajectories around the QCBH, focusing on how the quantum parameter affects photon trajectories near the event horizon, the shadow boundary, and the distribution of bright rings. We consider light rays originating from a static thin accretion disk in the equatorial plane and perform backward tracing from a distant observer. The photon trajectories are determined by Eq.~\eqref{trajectory_equation}. For convenience, we rewrite Eq.~\eqref{trajectory_equation} in the following form:
\begin{equation}
	\frac{{\rm d}\varphi}{{\rm d}r} = \pm \frac{b}{C(r)} \sqrt{\frac{1}{(1 - b^2 V_{\text{eff}}(r))B(r)}},
	\label{photon_trajectory_eq}
\end{equation}
where the sign choice depends on the photon's propagation direction during ray tracing. Specifically, for incoming light rays from the observer toward the BH, backward tracing begins with the negative sign when using numerical integration to plot the light rays. For photons with $b > b_{\rm c}$, they encounter a turning point at radial distance $r > r_{\text{ph}}$ where their radial velocity reverses. From this turning point, integration continues outward taking the positive sign in Eq.~\eqref{photon_trajectory_eq} toward the observer. For photons with $b < b_{\rm c}$, they no turning point occurs; instead, they cross the photon sphere and eventually reach the event horizon, with the entire integration performed using the negative sign. The critical case $b = b_{\rm c}$ corresponds to photons asymptotically approaching the unstable circular orbit at $r_{\rm ph}$.

For a given impact parameter $b$, azimuthal angle $\varphi$ is computed by integrating $\frac{{\rm d}\varphi}{{\rm d}r}$ along the photon trajectory. We then define the number of orbits as $n = \varphi/(2\pi)$. It is related to the number $m$ of intersections with the accretion disk through the expression:
\begin{equation}
	n(b) = \frac{2m - 1}{4}, \quad m = 1, 2, 3, \ldots.
	\label{nb_relation}
\end{equation}
Conversely, for a given $n$ (i.e., given $\varphi$), we can also solve Eq.~\eqref{photon_trajectory_eq} to find the two corresponding $b$ values, denoted by $b^\pm_m$, where $b^-_m < b_{\rm c}$ and $b^+_m > b_{\rm c}$. Then, the light trajectories can be classified in terms of the value of $n$ as three distinct types~\cite{Peng:2020wun,Yang:2022btw}:
\begin{itemize}
	\item \textbf{Direct:} $\frac{1}{4} < n <\frac{3}{4}$ corresponds to $m = 1$ and $b/M \in \left( b_{1}^{-}/M,b_{2}^{-}/M \right) \cup \left( b_{2}^{+}/M,\infty \right)$.
	\item \textbf{Lensed:} $\frac{3}{4}< n < \frac{5}{4}$ corresponds to $m = 2$ and $b/M \in \left( b_{2}^{-}/M,b_{3}^{-}/M \right) \cup \left( b_{3}^{+}/M,b_{2}^{+}/M \right)$.
	\item \textbf{Photon ring:} $n > \frac{5}{4}$ corresponds to $m \geq 3$ and $b/M \in \left( b_{3}^{-}/M,b_{3}^{+}/M \right)$.
\end{itemize}

In Fig.~\ref{fig:total_number_of_orbit}, we depicts the total number $n$ of orbits with $\zeta/M=1.5$. Meanwhile, we tabulate the values of $b^\pm_m$ and the other four physical quantities obtained in the previous section for different values of $\zeta/M$ in Table~\ref{table:1}.

\begin{figure}[htbp]
	\centering
	\includegraphics[width=7.5cm]{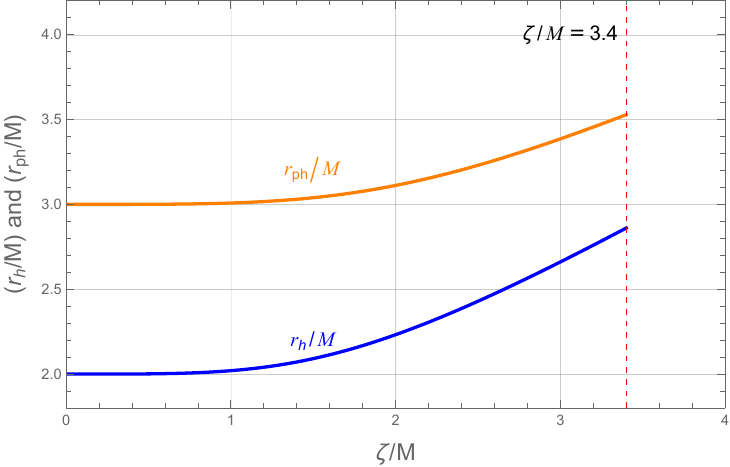}

	\caption{Relative positions of the $r_{\rm h}$ (blue) and $r_{\rm ph}$ (orange), with the vertical red dashed line marking the maximum allowed $\zeta/M$.}
	\label{fig:Model3_function_rh_and_rph_with_zeta}
\end{figure}

\begin{figure}[htbp]
	\centering
	\includegraphics[width=6.8cm]{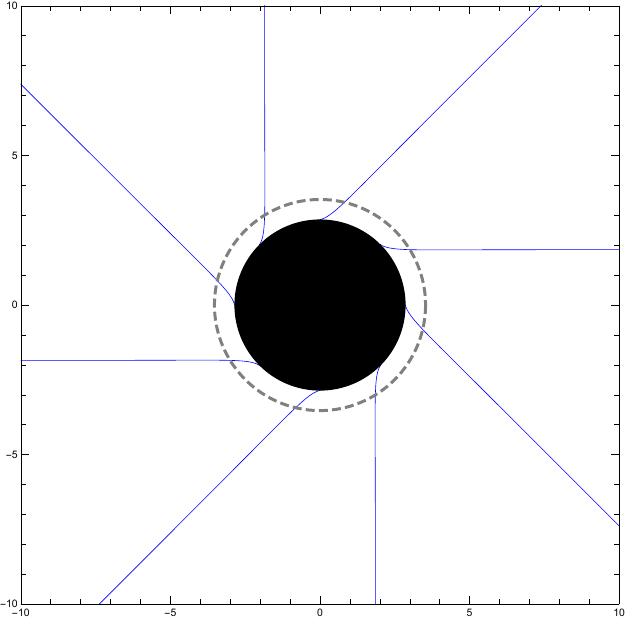}

	\caption{Photon trajectories from different directions for $\zeta/M=3.4$. The black disk shows the QCBH, and a dashed black curve marks the photon sphere.}
	\label{fig:photon_trajectory_from_all_direction}
\end{figure}

Finally, based on the above classification and using Eq.~\eqref{trajectory_equation}, we plot photon trajectories near the QCBH in Fig.~\ref{fig:photon_trajectories}. The image clearly illustrates that with increasing values of $\zeta/M$, both the photon sphere and the event horizon expand, and their separation decreases. We quantitatively show this change in Fig.~\ref{fig:Model3_function_rh_and_rph_with_zeta}. Meanwhile, as $\zeta/M$ increases, the light rays gradually exhibit special deflection around the event horizon of the QCBH, which is particularly evident in the image for $\zeta/M = 3.4$. Although Fig.~\ref{fig:photon_trajectories} suggests that only light rays on the right side of the event horizon are affected, spherical symmetry implies that this influence occurs in all directions. We plot eight photon trajectories that are symmetric about the origin in Fig.~\ref{fig:photon_trajectory_from_all_direction} to illustrate this clearly.

\section{Image of the quantum-corrected BH}\label{section4}

With the foundation laid in the preceding sections, we now turn to analyze the image of the QCBH. To simplify, we consider that an observer is positioned at the north pole and the disk emits radiation with a specific intensity $I^{\rm em}_{\nu}(r)$~\cite{Wang:2022yvi,Gralla:2019xty}. As discussed previously, the impact parameter $b$ of a photon determines the number of times its trajectory intersects the equatorial plane of the disk. Each such crossing at a radial distance $r$ contributes an additional term to the total intensity detected by the observer. Thus, the overall observed intensity is the superposition of all individual emission contributions, which can be expressed in the following form~\cite{Gralla:2020srx}:
\begin{equation}
	I_{\rm obs}(b) = \sum_{m} f_{m}A(r)^{2} I_{\rm em}(r) \bigg|_{r = r_m(b)}\label{total_observed_intensity}.
\end{equation}
Here $f_{m}$ is the fudge factor, whose value depends on the type of accretion disk. For thin disks, it typically takes~\cite{DeMartino:2023ovj}
\begin{equation}
	f_{m} =
		\begin{cases}
			1, & m = 1,2,3,\\
			0, & m > 3.
		\end{cases}\label{fudge_factor}
\end{equation}
And $I_{\rm em}(r) := \int I_{\nu}^{\rm em}(r) \rm d\nu$ represents the total emission profile at radius $r$ (integrated over all frequencies). The function $r_m(b)$ is called the transfer function. During backward ray-tracing, the transfer function gives the radial coordinate $r$ where a photon with an impact parameter $b$ intersects the disk for the $m$-th time. By selecting the value of $m$ and combining Eqs.~\eqref{photon_trajectory_eq} and \eqref{nb_relation}, the corresponding transfer functions can be obtained.

\begin{figure}[htb]
		\centering
	\includegraphics[width=7.5cm]{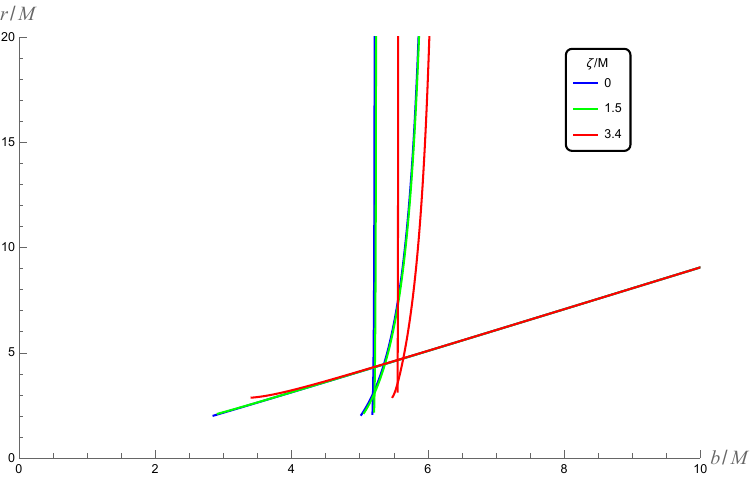}

	\caption{The transfer functions of the QCBH with $\zeta/M =1.5$ (green) and $\zeta/M =3.4$ (red) vs. Schwarzschild BH (blue)}
	\label{fig:transfer_functions}
\end{figure}

Figure~\ref{fig:transfer_functions} shows the first three transfer functions $r_m(b)$ for the Schwarzschild BH and QCBH. On one hand, the first three transfer functions $r_m(b)$ for the Schwarzschild BH and QCBH share some common characteristics. For $m=1$, the rate of change ${\rm d}r_{1}/{\rm d}b$ is very close to 1. This term mainly produces the direct image of the accretion disk, essentially reflecting the redshift effect on the source profile distribution $I_{\rm em}(r)$. For $m=2$, ${\rm d}r_{2}/{\rm d}b$ increases rapidly. It corresponds to light rays that intersect the accretion disk of the BH at most twice (lensed ring). These rays form a highly demagnified and faint image, confined to a much narrower range of $b$ compared to the direct image. For $m=3$, ${\rm d}r_{3}/{\rm d}b$ increases very quickly. It corresponds to light rays that intersect the accretion disk of the BH three times (photon ring). These rays form an extremely demagnified image confined to a very narrow range of $b$. Due to excessive demagnification, the images produced by higher-order transfer functions ($m > 3$) are negligible~\cite{Gralla:2019xty}. On the other hand, the transfer function graphs shift rightward and upward as the quantum parameter increases.
\begin{figure*}[htbp]
	\subfigure[Model-I]{\includegraphics[width=0.32\textwidth]{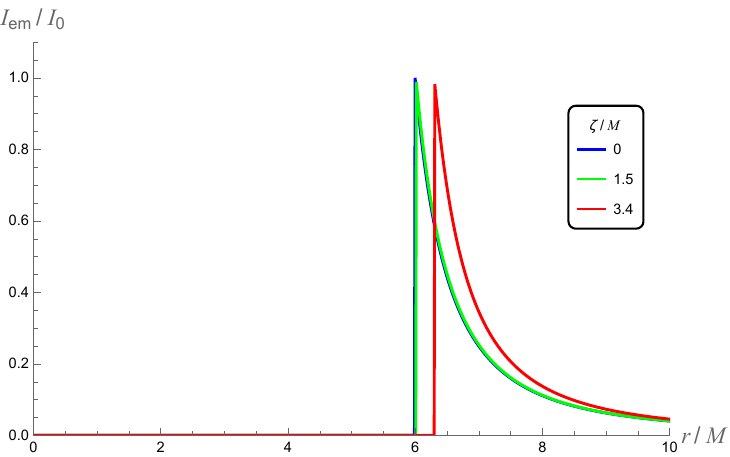}}
	\hfill
	\subfigure[Model-II]{\includegraphics[width=0.32\textwidth]{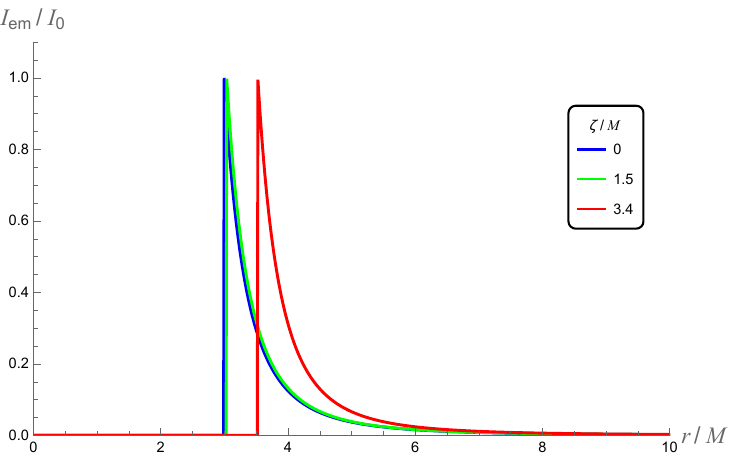}}
	\hfill
	\subfigure[Model-III]{\includegraphics[width=0.32\textwidth]{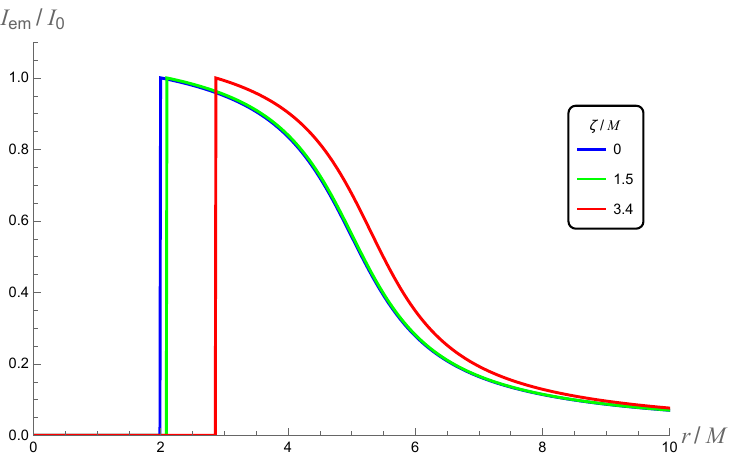}}

	\caption{Emission intensity profiles $I_{\rm em}(r)$ for the three models considered, shown for different values of $\zeta/M$.}
	\label{fig:Iem}
\end{figure*}

\begin{figure*}[htbp]
	\centering
	\subfigure[Model-I]{\includegraphics[height=4.5cm,keepaspectratio]{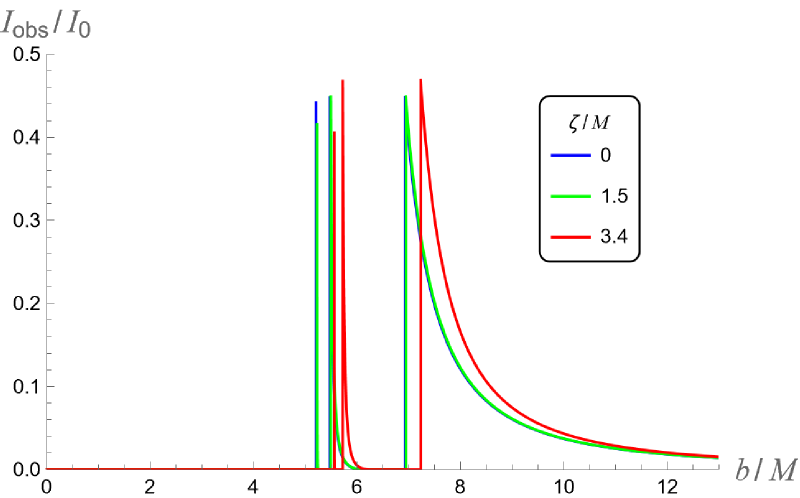}}
	\hspace{20pt}
	\subfigure[Optical-I.]{\includegraphics[height=4.5cm,keepaspectratio]{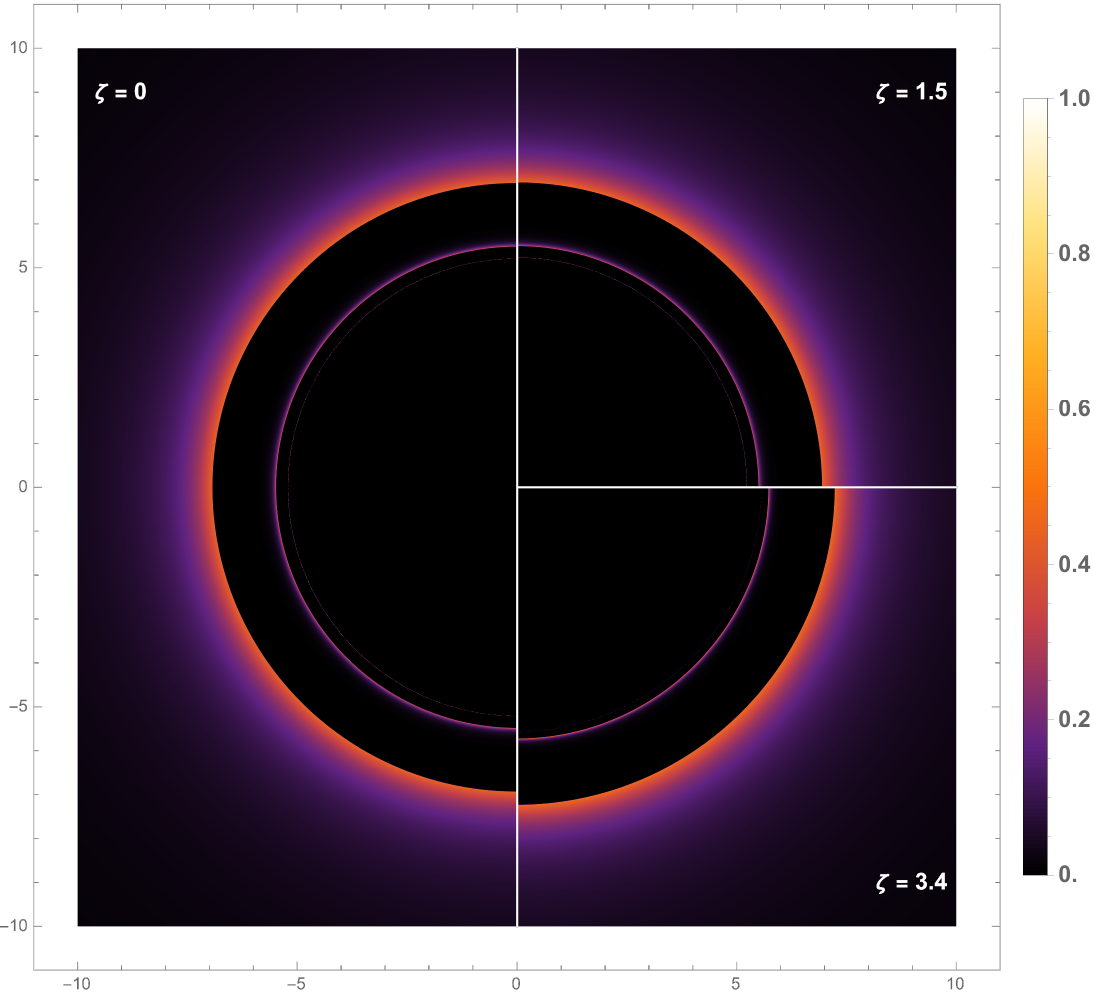}}
	\hfill

	\subfigure[Model-II]{\includegraphics[height=4.5cm,keepaspectratio]{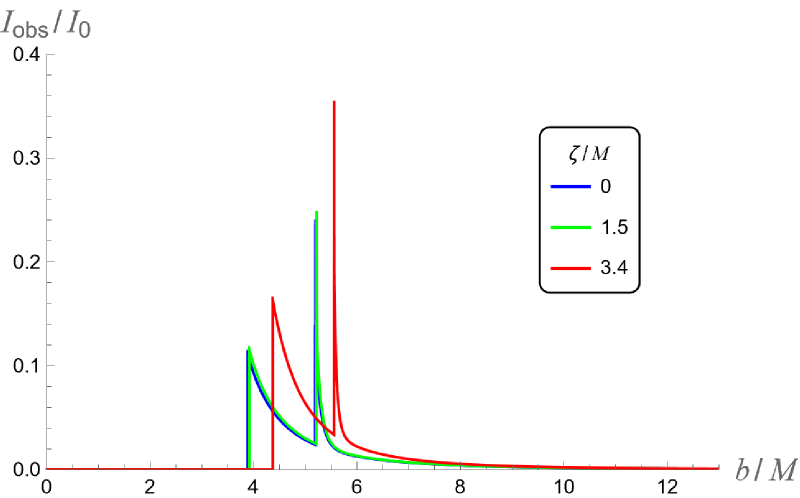}}
	\hspace{20pt}
	\subfigure[Optical-II.]{\includegraphics[height=4.5cm,keepaspectratio]{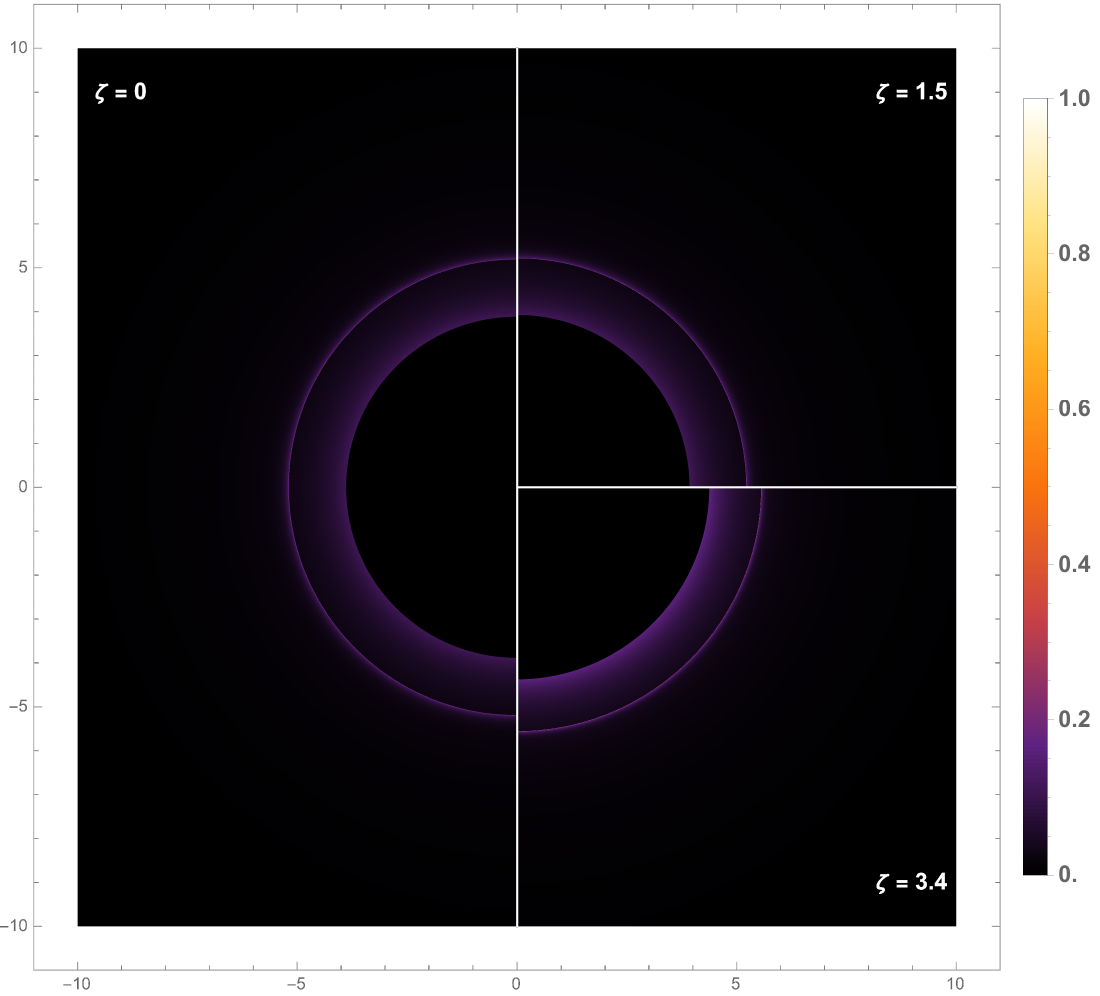}}
	\hfill

	\subfigure[Model-III]{\includegraphics[height=4.5cm,keepaspectratio]{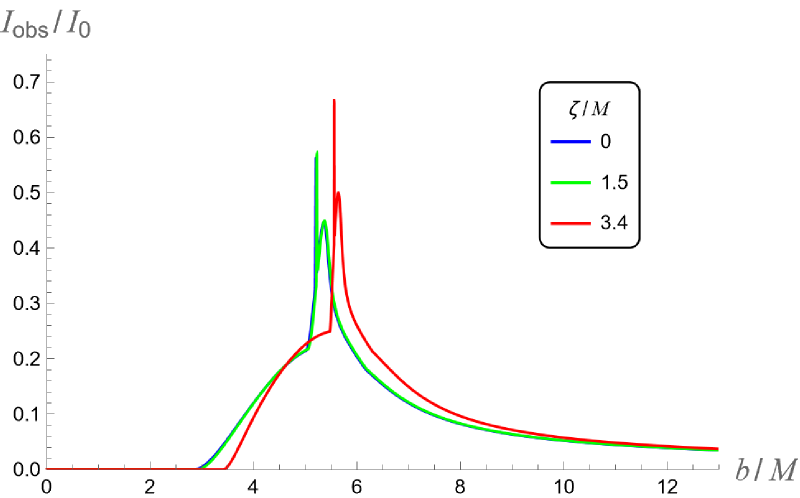}}
	\hspace{20pt}
	\subfigure[Optical-II.]{\includegraphics[height=4.5cm,keepaspectratio]{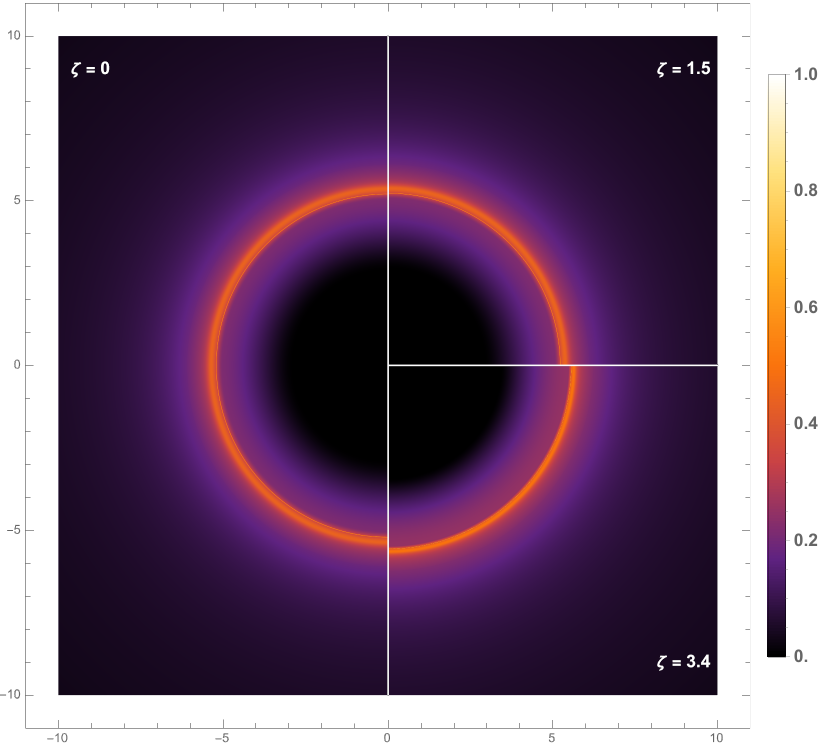}}

	\caption{The observed intensity (left) and the optical appearance (right) of three accretion disks around a QCBH under different $\zeta/M$ values. Each optical appearance image includes: Schwarzschild (left), $\zeta/M=1.5$ (upper right), and $\zeta/M=3.4$ (lower right).}
	\label{fig:Iobs_and_shadow}
\end{figure*}
\begin{figure*}[htbp]
	\centering
	\subfigure[Model-I]{\includegraphics[width=0.28\textwidth]{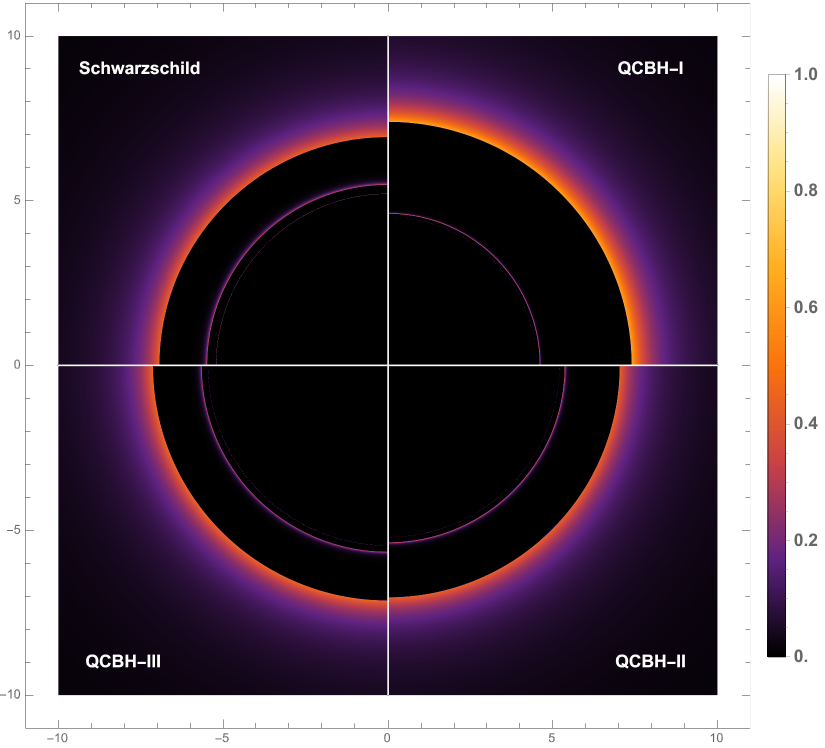}}
	\hspace{25pt}
	\subfigure[Model-II]{\includegraphics[width=0.28\textwidth]{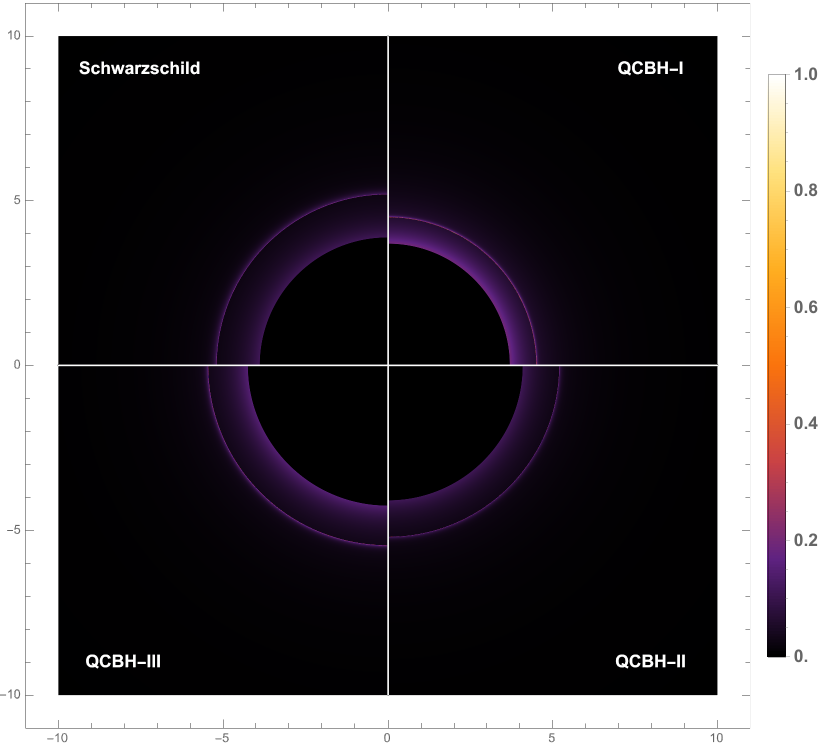}}
	\hspace{25pt}
	\subfigure[Model-III]{\includegraphics[width=0.28\textwidth]{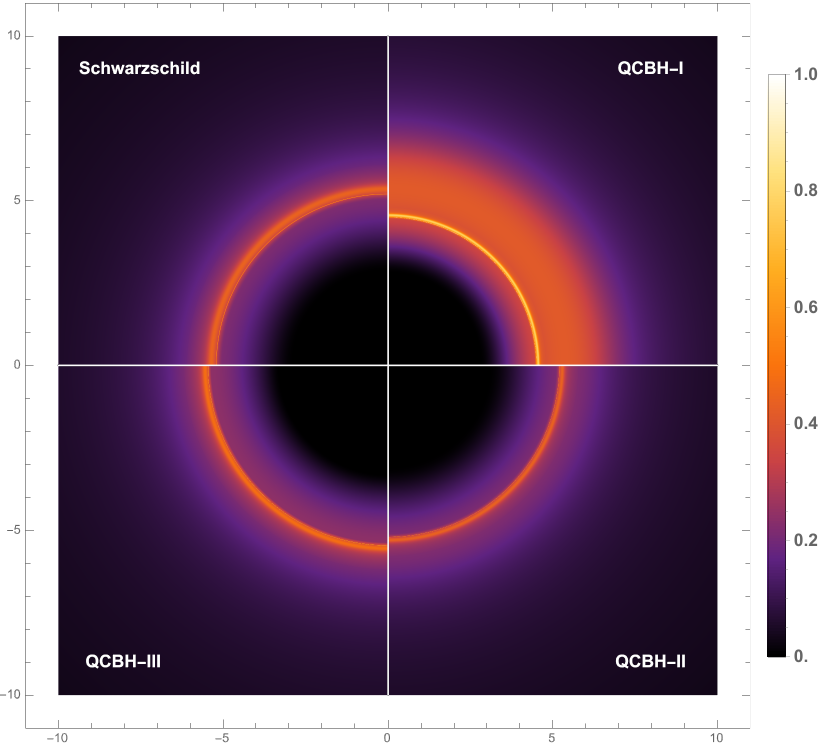}}

	\caption{Comparison of the optical appearances of the three QCBHs and the Schwarzschild BH under three simplified emission profiles when $\zeta/M=3.0$.}
	\label{comparison}
\end{figure*}

\subsection{Three simplified emission profiles}

It turns out that the observed image features, including the brightness patterns and the structure of photon rings, are strongly dependent on the choice of emission model. Therefore, it is crucial to precisely define the emission intensity profile $I_{\rm em}(r)$. To systematically investigate how the BH's optical appearance is influenced by the quantum parameter $\zeta/M$, we employ three simplified emission profiles, which are defined as follows~\cite{Wang:2022yvi,Gralla:2019xty}:
\begin{itemize}
	\item Model-I: Emission peaks at the $r_{\rm isco}$ and decays rapidly outward:
	\begin{equation}
		I_{\rm em}(r) :=
		\begin{cases}
			I_0 \left[ \dfrac{1}{r - (r_{\rm isco} - 1)} \right]^2, & r > r_{\rm isco}, \\
			0, & r \leq r_{\rm isco}.
		\end{cases}\label{first_emission_intensity}
	\end{equation}

	\item Model- II: Emission peaks at the $r_{\rm ph}$ and decays rapidly outward:
	\begin{equation}
		I_{\rm em}(r) :=
		\begin{cases}
			I_0 \left[ \dfrac{1}{r - (r_{\rm ph} - 1)} \right]^3, & r > r_{\rm ph}, \\
			0, & r \leq r_{\rm ph}.
		\end{cases}
	\end{equation}\label{second_emission_intensity}

	\item Model-III: Emission starts just outside the $r_{\rm h}$ and decays slowly outward:
	\begin{equation}
		I_{\text{em}}(r) :=
		\begin{cases}
			I_0 \dfrac{\frac{\pi}{2} - \arctan[r - (r_{\text{isco}} - 1)]}{\frac{\pi}{2} - \arctan[r_{\rm h} - (r_{\rm isco} - 1)]}, & r > r_{\rm h}, \\
			0, & r \leq r_{\rm h}.
		\end{cases}\label{third_emission_intensity}
	\end{equation}
\end{itemize}
Here $I_{0}$ represents the peak emission intensity. Figure~\ref{fig:Iem} illustrates the emission intensities $I_{\rm em}/I_0$ for the three different emission models. As $\zeta/M$ increases, the peak positions of the three models gradually shift outward.

Furthermore, by using Eq.~\eqref{total_observed_intensity}, we compute the observed intensity distributions for the QCBH under the three emission models and plot the corresponding optical appearances; the results are shown in Fig.~\ref{fig:Iobs_and_shadow}. These results demonstrate that the quantum parameter $\zeta/M$ will lead to the observable characteristics. Specifically, as $\zeta/M$ increases, the shadow of the QCBH becomes more extensive compared to that of the Schwarzschild BH. Simultaneously, the growth of $\zeta/M$ compresses the lensed and photon ring structures. For distinguishable rings, the interval between their received intensity peaks further decreases with the increase of $\zeta/M$. In contrast, previous studies on QCBH-I and QCBH-II revealed different behavioral patterns: as $\zeta/M$ increases, the central dark region of QCBH-I shrinks while its optical appearance becomes brighter, whereas QCBH-II shows almost no alteration in its optical appearance~\cite{Chen:2025ifv}. To more clearly illustrate the distinct influences of the quantum parameter on the optical appearances of these three QCBHs, we plot their optical appearances for $\zeta/M=3.0$ and compare with that of the Schwarzschild BH in Fig.~\ref{comparison}, where the QCBH studied in this paper is temporarily denoted as “QCBH-III”. Therefore, by systematically examining the shadow size and the luminosity distribution characteristics of the bright rings, one can not only effectively differentiate the QCBH from the Schwarzschild case but also accurately identify specific quantum-corrected models, thereby providing a more precise research pathway for testing quantum gravity in strong-field environments.

\subsection{Emission profiles with Johnson's standard unbound distribution}

\begin{table}[htbp]
	\centering

	\caption{Parameter selection for different SU emission models.}
	\setlength{\tabcolsep}{9pt}
	\begin{tabular}{cccc}
		\hline
		Modle&$\mu$&$\gamma$&$\delta/M$\\
		\hline
		SU1 & 17$r_{\rm h}$/6 & $-$2 & 1/4 \\
		SU2 & 2$r_{\rm h}$ & $-$2 & 3/2 \\
		SU3 & 3$r_{\rm h}$/2 & 2 & 1/4 \\
		SU4 & $r_{\rm h}$ & 0 & 1 \\
		SU5 & 0 & $-$2 & 3/2
		\label{table:2}\\
		\hline
	\end{tabular}
\end{table}
\begin{figure}[hbp]
	\centering
	\includegraphics[width=8cm]{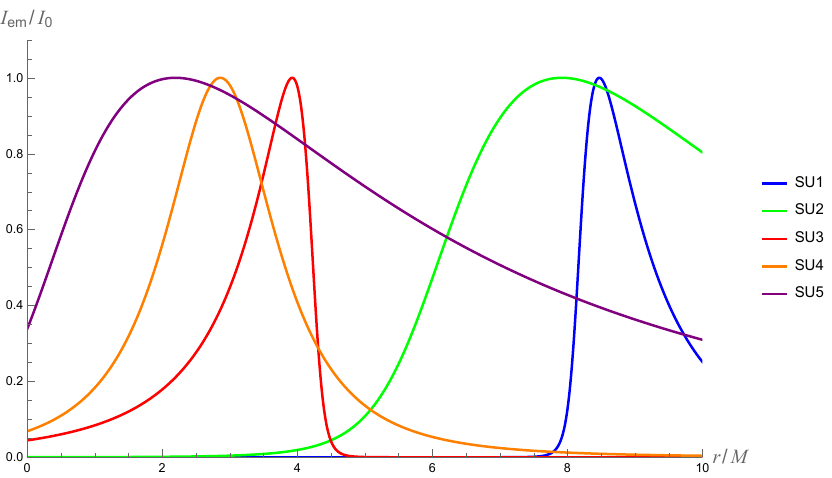}

	\caption{Normalized emission intensity of five SU models for the QCBH, at $\zeta/M=3.4$.}
	\label{fig:SU_Iem}
\end{figure}
\begin{figure*}[hb]
 \begin{minipage}{0.05\textwidth}
	\includegraphics[height=2.4cm]{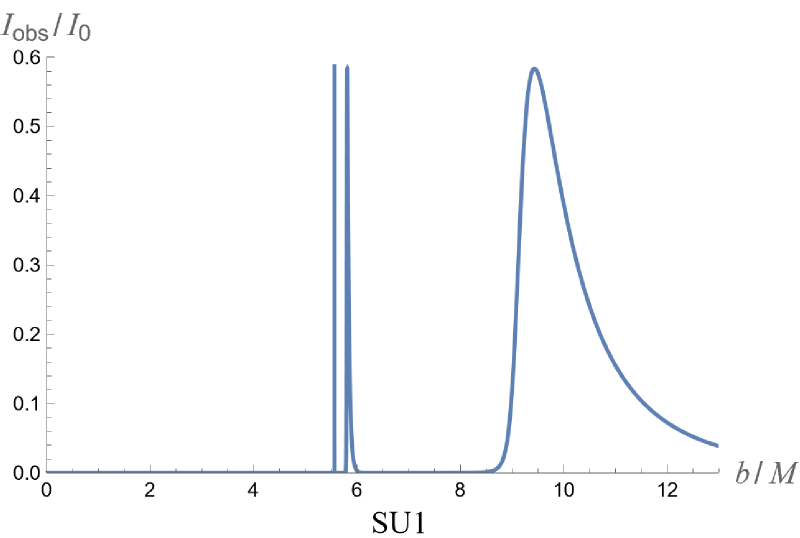}
 \end{minipage}
 \hfill
 \begin{minipage}{0.15\textwidth}
	\includegraphics[height=2.4cm,keepaspectratio]{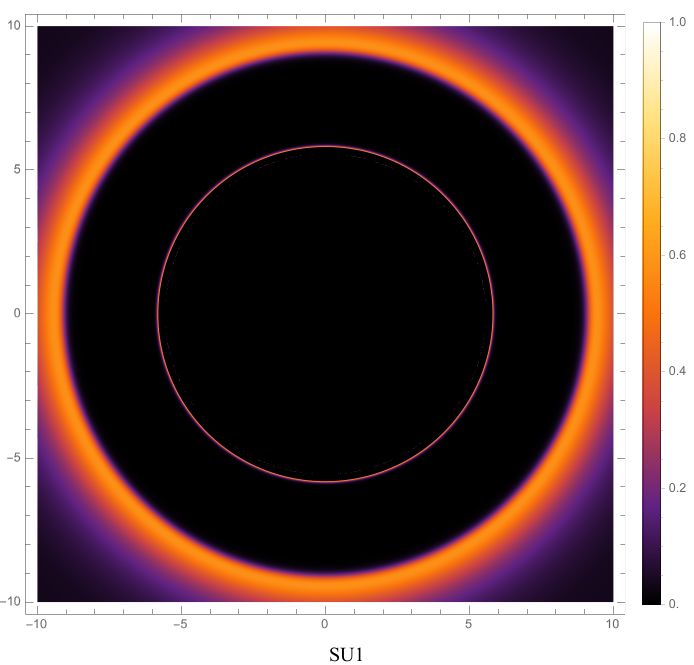}
 \end{minipage}
 \hfill
 \begin{minipage}{0.05\textwidth}
	\includegraphics[height=2.4cm,keepaspectratio]{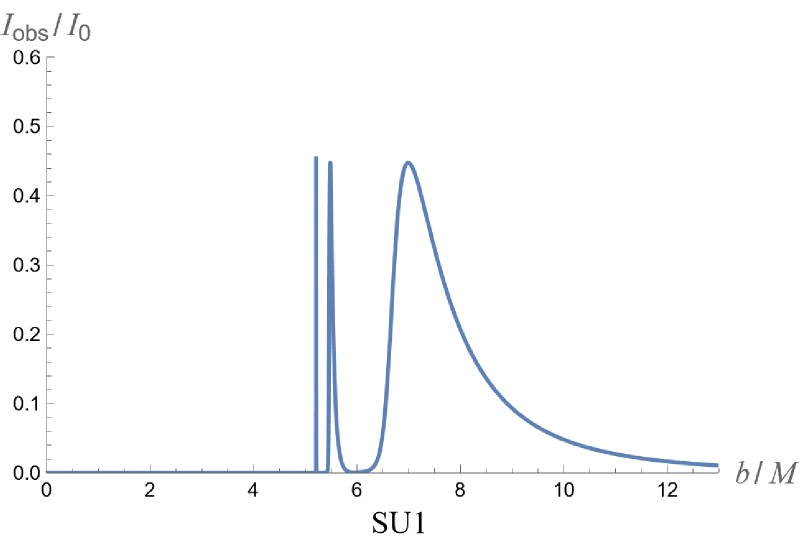}
 \end{minipage}
 \hfill
 \begin{minipage}{0.15\textwidth}
	\includegraphics[height=2.4cm]{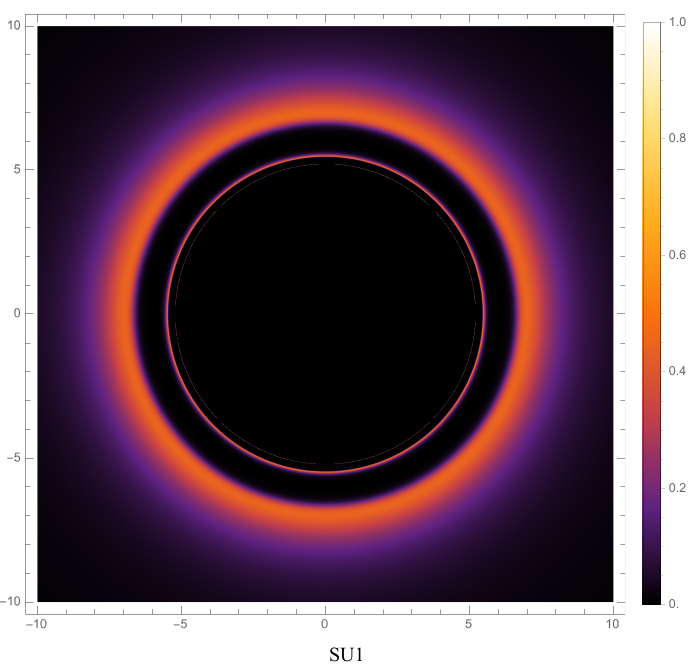}
 \end{minipage}
 \hfill

 \begin{minipage}{0.05\textwidth}
	\includegraphics[height=2.4cm]{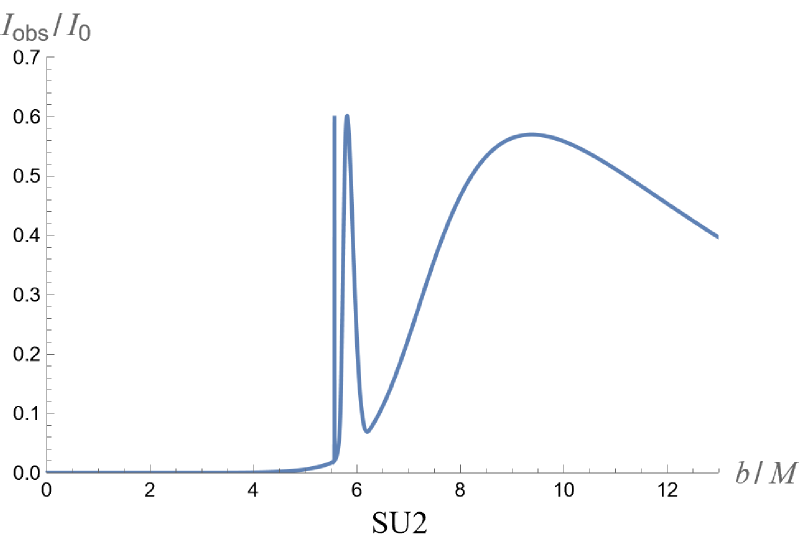}
 \end{minipage}
 \hfill
 \begin{minipage}{0.15\textwidth}
	\includegraphics[height=2.4cm,keepaspectratio]{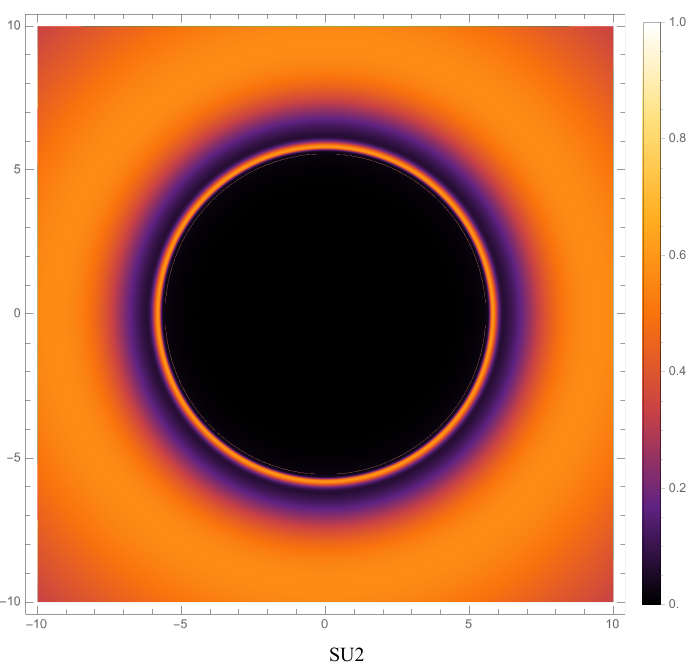}
 \end{minipage}
 \hfill
 \begin{minipage}{0.05\textwidth}
	\includegraphics[height=2.4cm,keepaspectratio]{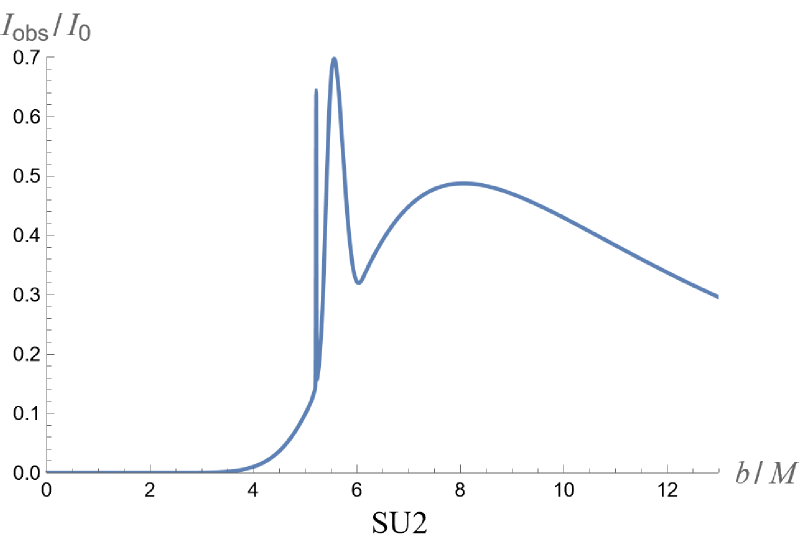}
 \end{minipage}
 \hfill
 \begin{minipage}{0.15\textwidth}
	\includegraphics[height=2.4cm]{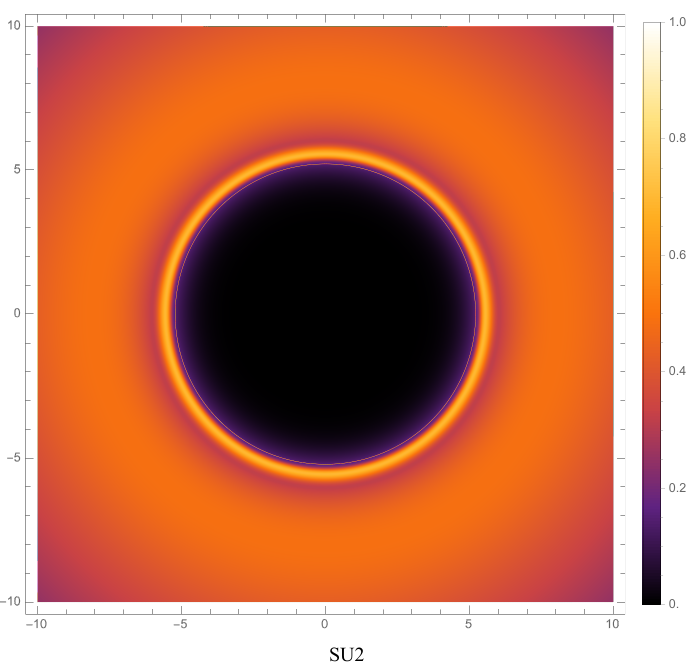}
 \end{minipage}
 \hfill

 \begin{minipage}{0.05\textwidth}
	\includegraphics[height=2.4cm]{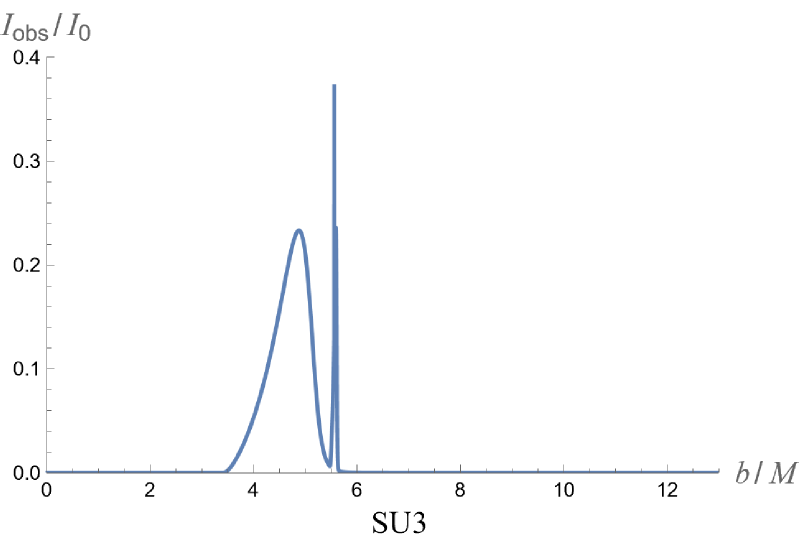}
 \end{minipage}
 \hfill
 \begin{minipage}{0.15\textwidth}
	\includegraphics[height=2.4cm,keepaspectratio]{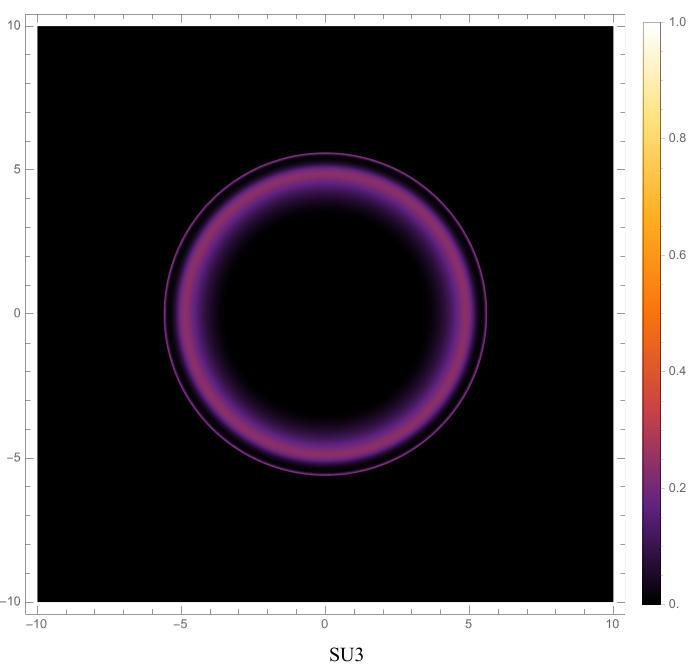}
 \end{minipage}
 \hfill
 \begin{minipage}{0.05\textwidth}
	\includegraphics[height=2.4cm,keepaspectratio]{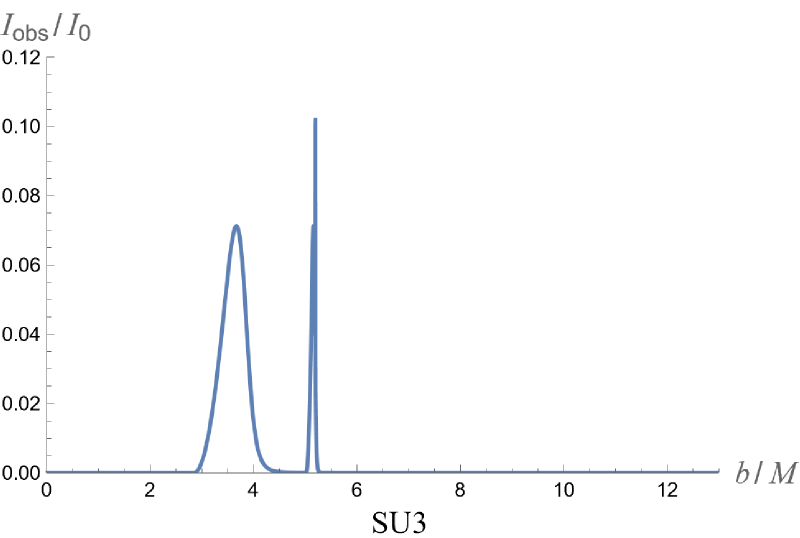}
 \end{minipage}
 \hfill
 \begin{minipage}{0.15\textwidth}
	\includegraphics[height=2.4cm]{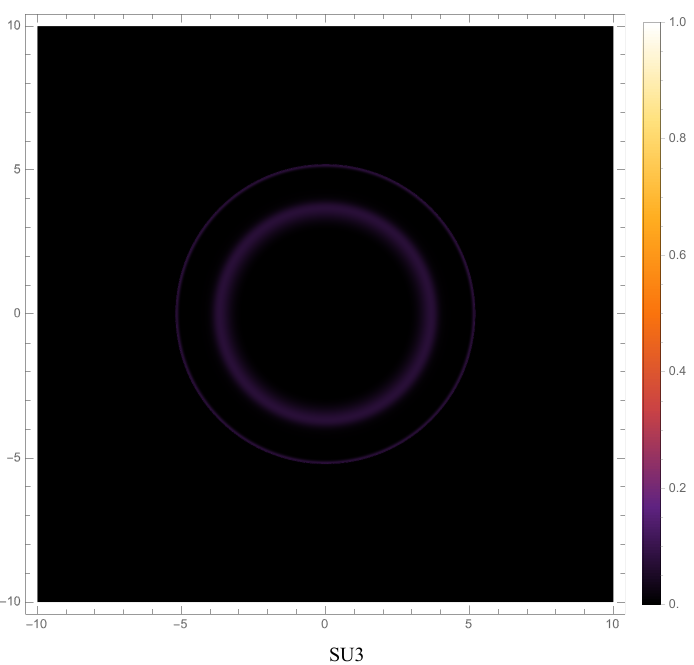}
 \end{minipage}
 \hfill

 \begin{minipage}{0.05\textwidth}
	\includegraphics[height=2.4cm]{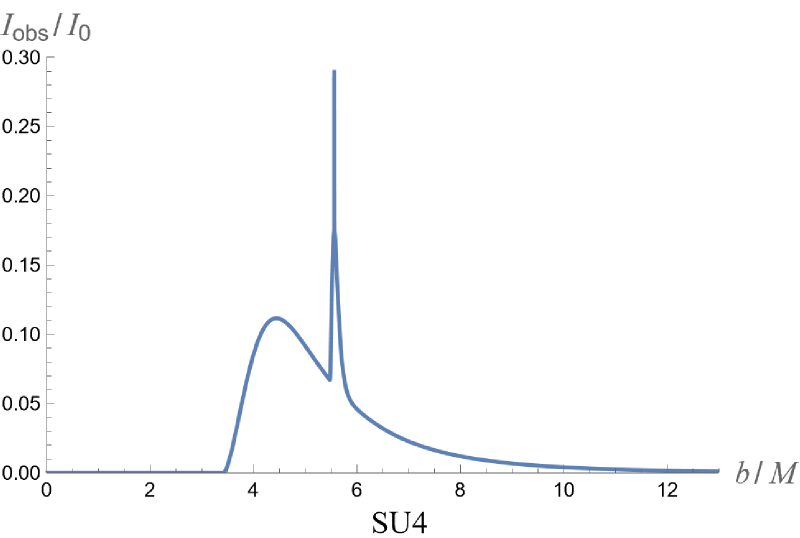}
 \end{minipage}
 \hfill
 \begin{minipage}{0.15\textwidth}
	\includegraphics[height=2.4cm,keepaspectratio]{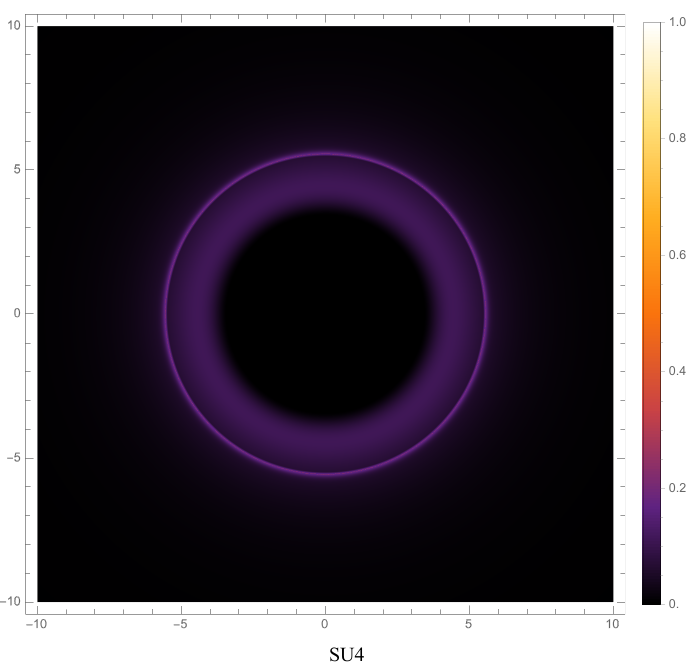}
 \end{minipage}
 \hfill
 \begin{minipage}{0.05\textwidth}
	\includegraphics[height=2.4cm,keepaspectratio]{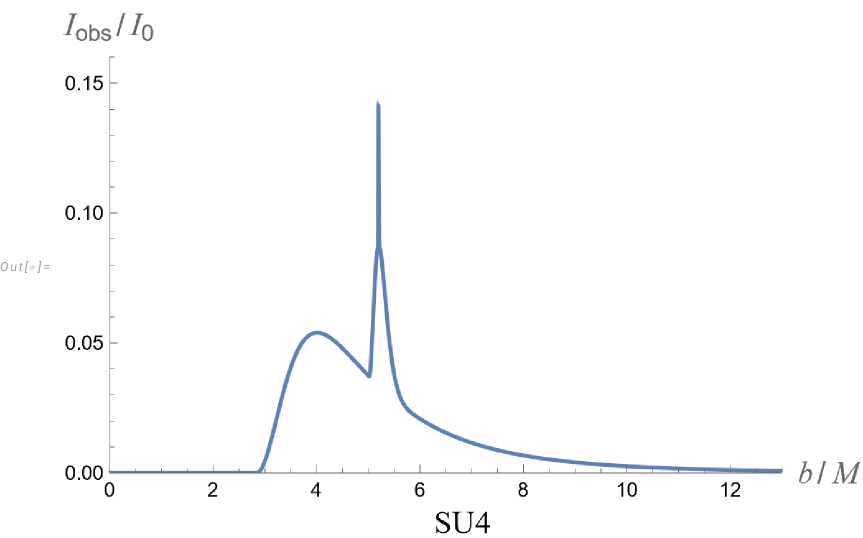}
 \end{minipage}
 \hfill
 \begin{minipage}{0.15\textwidth}
	\includegraphics[height=2.4cm]{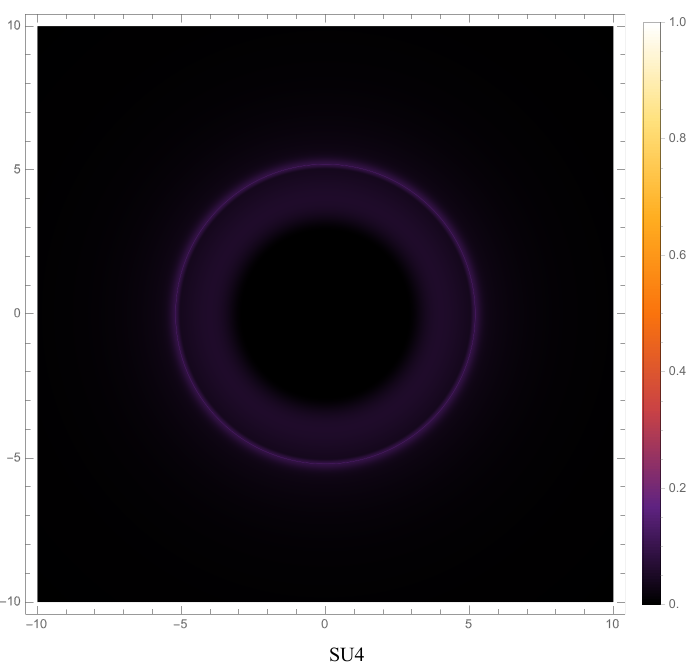}
 \end{minipage}
 \hfill

 \begin{minipage}{0.05\textwidth}
	\includegraphics[height=2.4cm]{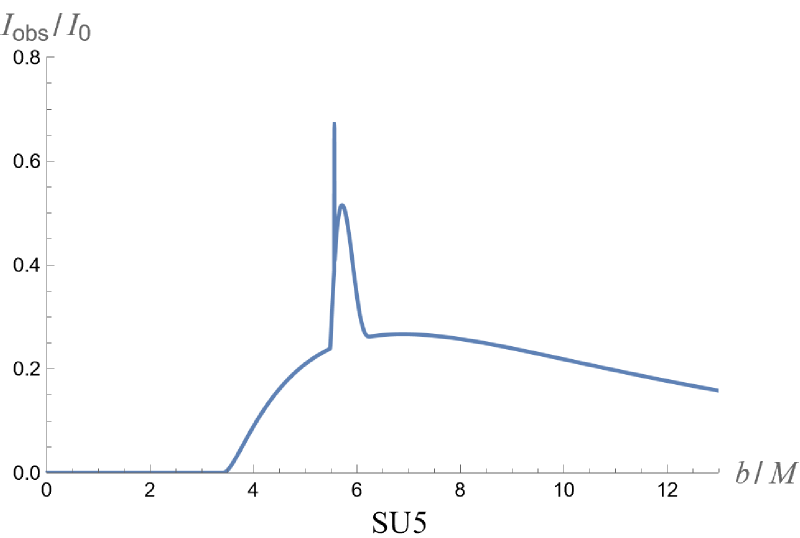}
 \end{minipage}
 \hfill
 \begin{minipage}{0.15\textwidth}
	\includegraphics[height=2.4cm,keepaspectratio]{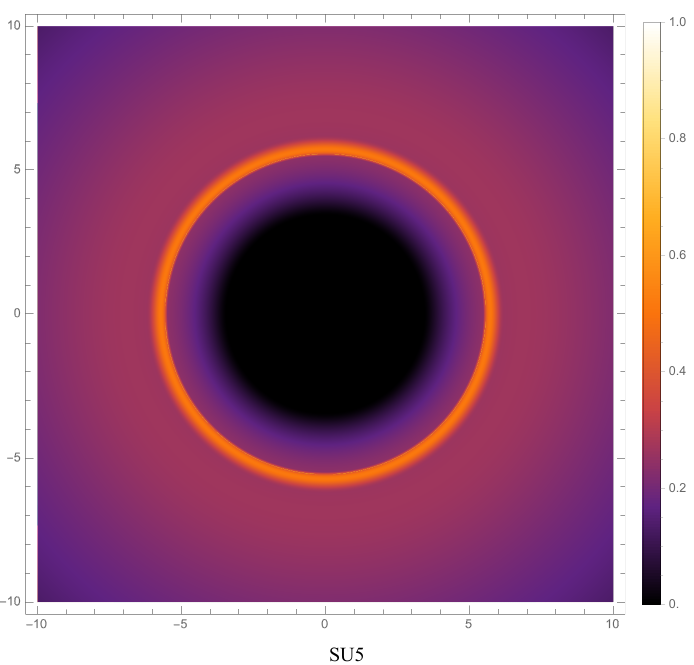}
 \end{minipage}
 \hfill
 \begin{minipage}{0.05\textwidth}
	\includegraphics[height=2.4cm,keepaspectratio]{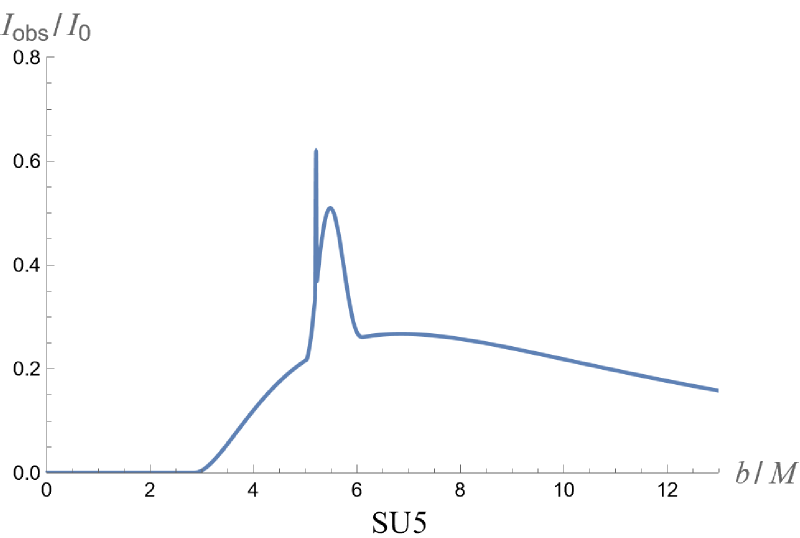}
 \end{minipage}
 \hfill
 \begin{minipage}{0.15\textwidth}
	\includegraphics[height=2.4cm]{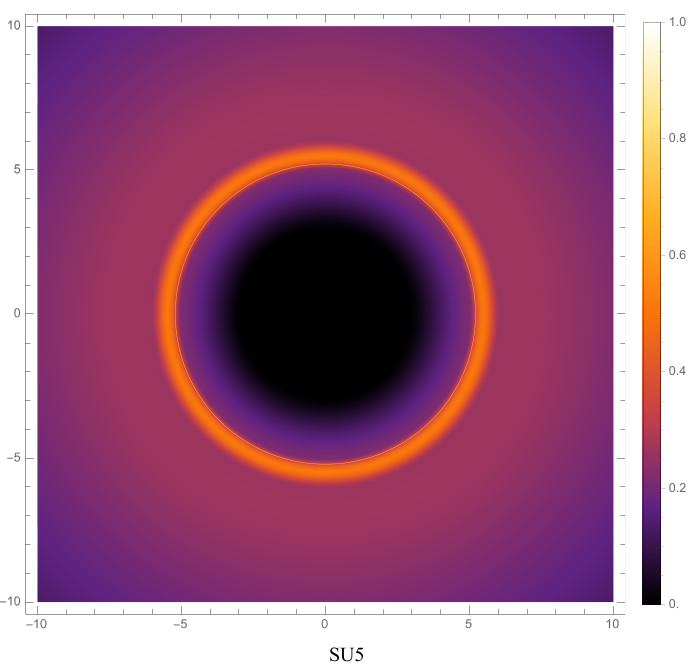}
 \end{minipage}
 \hfill

	\caption{Observed intensity and optical appearance for five SU emission models. Left panels: QCBH ($\zeta/M=3.4$). Right panels: Schwarzschild BH.}
	\label{fig:SU_Iobs_and_shadow}
\end{figure*}

\begin{figure*}[htb]
	\centering
	\subfigure[SU1]{\includegraphics[width=0.32\textwidth]{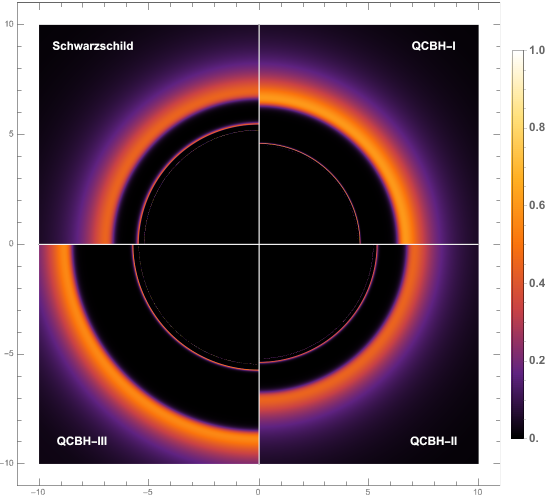}}
 \hfill
 \subfigure[SU2]{\includegraphics[width=0.32\textwidth]{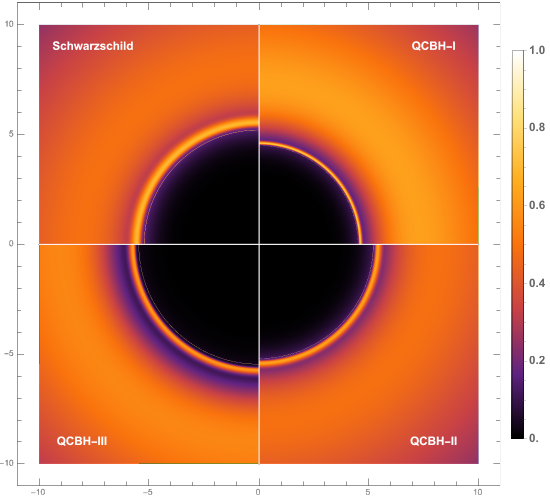}}
 \hfill
 \subfigure[SU3]{\includegraphics[width=0.32\textwidth]{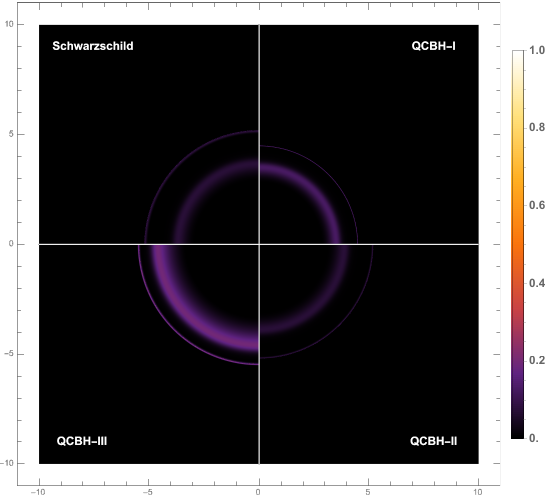}}

 \subfigure[SU4]{\includegraphics[width=0.32\textwidth]{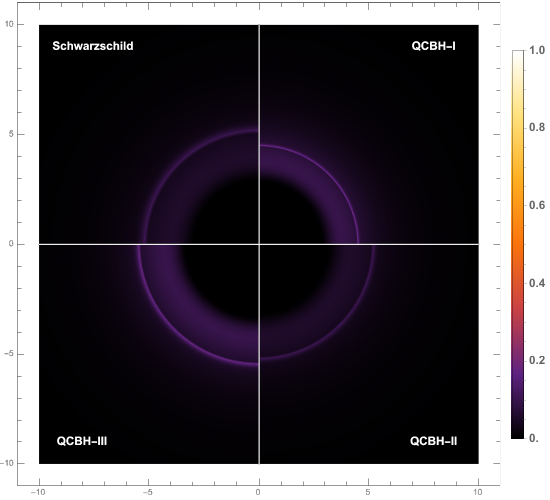}}
 \subfigure[SU5]{\includegraphics[width=0.32\textwidth]{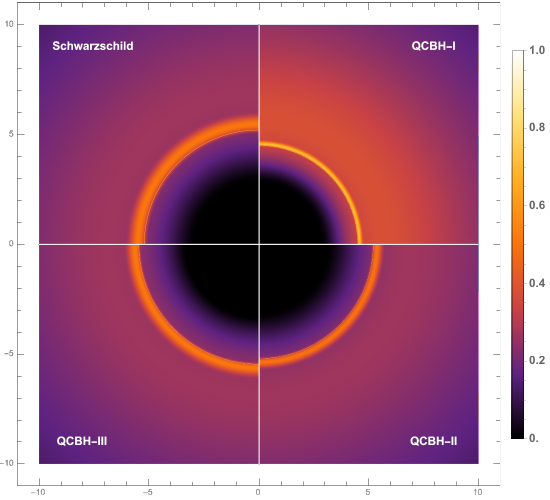}}

	\caption{Comparison of the optical appearances of the three QCBHs and the Schwarzschild BH under five SU emission models when $\zeta/M=3.0$.}
	\label{comparisonSU}
\end{figure*}

To extend our investigation beyond the simplified emission models, we adopt Johnson's standard unbound (SU) distribution as a more flexible profile for accretion disk simulations, which takes the following functional form~\cite{Gralla:2020srx}:
\begin{equation}
	I_{\rm em}(r;\gamma,\mu,\delta) = \frac{e^{-\frac{1}{2}\left[\gamma + \rm{arcsinh}\left(\frac{r - \mu}{\delta}\right)\right]^{2}}}{\sqrt{(r - \mu)^{2} + \delta^{2}}}.
	\label{eq:SU_distribution}
\end{equation}
This emission models enables the modeling of a broad spectrum of emission characteristics by varying key parameters that control the approximate peak location ($\mu$), profile asymmetry ($\gamma$), and radial extent ($\delta$) of the intensity. Following the reference values for parameter settings provided in Ref.~\cite{DeMartino:2023ovj}, we select five emission models. The corresponding parameter values are presented in Table~\ref{table:2}. For convenience of presentation, we name these models SU1 to SU5 in descending order of the $\mu$ values. To clearly demonstrate the effects of quantum-corrections in this QCBH, we select $\zeta/M=3.4$ and display the function graphs of these five emission models in Fig.~\ref{fig:SU_Iem}. Furthermore, we present the observed intensity and optical appearance for the QCBH and Schwarzschild BH under their respective SU models in Fig.~\ref{fig:SU_Iobs_and_shadow}. The optical appearance consistently concurs with the conclusions obtained from the simulations of the three simplified emission models. Namely, the QCBH possesses a larger shadow, along with narrower lensed and photon rings and reduced spacing between them. Similarly, we plot a comparison of the optical appearances of the three QCBHs and the Schwarzschild BH under our selected five emission models for $\zeta/M=3.0$ in Fig.~\ref{comparisonSU}. It is worth noting that although the optical appearances of QCBH-I and QCBH-II under the SU emission model were not studied in~\cite{Chen:2025ifv}, it is not difficult to see from the figure that the conclusions drawn there still hold under this emission model, and the quantum parameter exerts distinctly different influences on the optical appearance across different QCBHs.

\section{Summary}\label{section5}

This paper presents a systematic study of the optical appearance of a QCBH without Cauchy horizons when illuminated by a static thin accretion disk. We first analyzed the influence of the quantum parameter on the four physical quantities ($r_{\rm h}$, $b_{\rm c}$, $r_{\rm ph}$, $r_{\rm isco}$), we found that all these quantities increase with $\zeta/M$. We then constrained the parameter $\zeta/M$ from both theoretical and observational perspectives. Theoretically, to ensure the existence of an event horizon, $\zeta/M$ must be less than $3.9374$; observationally, using EHT data on Sgr A*, we placed an observational constraint on $\zeta/M$ to be less than $3.4081$ at the $2\sigma$ confidence level. Ultimately, we adopted $\zeta/M \leq 3.4081$. Within the allowed parameter range, we further analyzed the trajectories of photons near the QCBH, and found that as $\zeta/M$ increases, photon trajectories exhibit slight deflection near the event horizon as shown in Figs.~\ref{fig:photon_trajectories} and \ref{fig:photon_trajectory_from_all_direction}. This phenomenon arises from a quantum-induced modification of the geodesic equations.

Subsequently, we systematically simulated the image of the QCBH under three typical emission profiles, and presented the results in Fig.~\ref{fig:Iobs_and_shadow}. We found that a large $\zeta/M$ enlarges the shadow region, along with narrower lensed and photon rings and reduced spacing between them. Moreover, we also simulated the image of the QCBH under large quantum parameters using the SU distribution (as shown in Fig.~\ref{fig:SU_Iobs_and_shadow}) and reached the same conclusion. As can be seen from the comparison between Fig.~\ref{comparison} and Fig.~\ref{comparisonSU}, the optical appearance features of the QCBH (such as shadow size, lensed ring, and photon ring) exhibit different trends with the quantum parameter compared to the two types of QCBHs studied in Ref.~\cite{Chen:2025ifv}.

In summary, the QCBH studied in this work displays observable features that are expected to be testable through future observations, as its optical appearance exhibits sufficiently significant differences from those of the Schwarzschild BH and the other two QCBHs (QCBH-I and QCBH-II). These distinctive signatures not only enable clear phenomenological discrimination but also contribute a crucial piece to the physical picture of the QCBH family. Additionally, images generated by incorporating rotation and more realistic astrophysical environments (such as thick accretion disks, inclination angles, etc.) will be more convincing, but this simplified study still provides a benchmark for future investigations into the appearance of QCBHs.

%-------------------------------------------------------------

\section*{Acknowledgment}

This work is supported in part by NSFC Grants No. 12165005 and No. 11961131013.

%%=============================================================================================

%%=============================================================================================

\end{document}